\documentclass{article}

\usepackage{arxiv}

\usepackage[utf8]{inputenc} % allow utf-8 input
\usepackage[T1]{fontenc}    % use 8-bit T1 fonts
\usepackage{hyperref}       % hyperlinks

\usepackage{url}            % simple URL typesetting
\usepackage{booktabs}       % professional-quality tables
\usepackage{amsfonts}       % blackboard math symbols
\usepackage{nicefrac}       % compact symbols for 1/2, etc.
\usepackage{microtype}      % microtypography
\usepackage{lipsum}		% Can be removed after putting your text content
\usepackage{doi}
\usepackage{epsfig,endnotes}
\usepackage{amssymb}
\usepackage{times}
\usepackage{algorithm}
\usepackage{algorithmicx}
\usepackage{algpseudocode}
\usepackage{amsmath}
\usepackage{graphicx,float}
\usepackage{multirow}
\usepackage{graphics}
\usepackage{comment}
\usepackage{listings}
\usepackage{enumitem}

\newcommand{\Sysname} {BW-Raft~}
\newcommand{\gr}{BW-Raft~}

\newcommand{\SEC}[1] {Section \ref{#1}}

\begin{document}

\title{Cost-effective BlackWater Raft on Highly Unreliable Nodes at Scale Out}

%\date{September 9, 1985}	% Here you can change the date presented in the paper title
%\date{} 					% Or removing it

\author{
Zichen Xu \#, Yunxiao Du\thanks{Du Yunxiao, Zhang Kanqi, Huang Jiacheng and Liu Jie participated in this project as undergraduates.} \#,  Kanqi Zhang \#, Jiacheng Huang \#, Jie Liu \#, Jingxiong Gao \#, Christopher Stewart $\dagger$ \\
\# Nanchang University, $\dagger$ Ohio State University\\
\# \texttt{xuz@ncu.edu.cn}, $\dagger$ \texttt{cstewart@cse.ohio-state.edu}\\
\# \texttt{\{yunxiaodu, zhangkanqi, jiachenghuang, liujie, jingxionggao\}@email.ncu.edu.cn } 
}
% Zichen Xu \#, Yunxiao Du $\dagger$   ,  Kanqi Zhang \#, Christopher Stewart *, Jiacheng Huang \# \\ %
%	  	
%	  	 \#xuz@ncu.edu.cn , $\dagger$yunxiao.du@zju.edu.cn ,\\    
%	 	\{zhangkanqi, jiachenghuang \}@email.ncu.edu.cn \#,
%	 * cstewart@cse.ohio-state.edu

%\hspace{1mm}David S.~Hippocampus \\
%	Department of Computer Science\\
%	Cranberry-Lemon University\\
%	Pittsburgh, PA 15213 \\
%	\texttt{hippo@cs.cranberry-lemon.edu} \\
	%% examples of more authors
%	\And
%	{\includegraphics[scale=0.06]{orcid.pdf}\hspace{1mm}Elias D.~Striatum} \\
%	Department of Electrical Engineering\\
%	Mount-Sheikh University\\
%	Santa Narimana, Levand \\
%	\texttt{stariate@ee.mount-sheikh.edu} \\
	%% \AND
	%% Coauthor \\
	%% Affiliation \\
	%% Address \\
	%% \texttt{email} \\
	%% \And
	%% Coauthor \\
	%% Affiliation \\
	%% Address \\
	%% \texttt{email} \\
	%% \And
	%% Coauthor \\
	%% Affiliation \\
	%% Address \\
	%% \texttt{email} \\

% Uncomment to remove the date
\date{}

% Uncomment to override  the `A preprint' in the header
\renewcommand{\headeright}{}
\renewcommand{\undertitle}{}
%\renewcommand{\shorttitle}{\textit{arXiv} Template}

%%% Add PDF metadata to help others organize their library
%%% Once the PDF is generated, you can check the metadata with
%%% $ pdfinfo template.pdf
%%%\hypersetup{
%%%pdftitle={A template for the arxiv style},
%%%pdfsubject={q-bio.NC, q-bio.QM},
%%%pdfauthor={David S.~Hippocampus, Elias D.~Striatum},
%%%pdfkeywords={First keyword, Second keyword, More},
%%%}

\maketitle
\begin{abstract}
The Raft algorithm maintains strong consistency across data replicas in Cloud.
This algorithm divides nodes into leaders and followers, to satisfy read/write requests spanning geo-diverse sites.
With the increase of workload, Raft shall provide scale-out performance in proportion.
However, traditional scale-out techniques encounter bottlenecks in Raft, and when the provisioned sites exhaust local resources, the performance loss will grow exponentially.
To provide scalability in Raft, this paper proposes a cost-effective mechanism for elastic auto-scaling in Raft, called BlackWater-Raft or BW-Raft.
BW-Raft extends the original Raft with the following abstractions: (1) \emph{secretary} nodes that take over expensive log synchronization operations from the leader, relaxing the performance constraints on locks. (2) massive low cost \emph{observer} nodes that handle reads only, improving throughput for typical data intensive services.
These abstractions are stateless, allowing elastic scale-out on unreliable yet cheap spot instances.
In theory, we demonstrate that BW-Raft can maintain Raft's strong consistency guarantees when scaling out, processing a 50X increase in the number of nodes compared to the original Raft.
We have prototyped the BW-Raft on key-value services and evaluated it with many state-of-the-arts on Amazon EC2 and Alibaba Cloud.
Our results show that within the same budget, BW-Raft's resource footprint increments are 5-7X  smaller than Multi-Raft, and 2X better than original Raft. 
Using spot instances, BW-Raft can reduces costs by 84.5\% compared to Multi-Raft.
In the real world experiments, BW-Raft improves goodput of the 95th-percentile SLO by 9.4X, thus serving as an alternative for services scaling out with strong consistency.     
\end{abstract}


\section{Introduction}\label{sec:INT}

Raft algorithm is invented to support strong consistency
in networked services, serving as a simplified alternative for Paxos~\cite{lamport2001paxos}.  Distributed systems use Raft to
keep consensus between software/system components.
For example, components must agree on locking
ownership and operations in synchronized queues.
Raft is used in practice by many large-scale platforms,
such as Google Kubernetes~\cite{kubernetes}, Core OS~\cite{coreos}, and Oracle~\cite{oracle}.
These platforms, i.e., the clients of Raft,
suffer inflated costs when Raft employs
inefficient scaling techniques.

%By design, Raft is easy to learn and implement.
Software threads in Raft are either {\em leaders}
or {\em followers}.  At all times, Raft allows zero or
one leader elected by followers.  Followers
periodically heartbeat the leader, checking for
failures.  Upon leader failure, followers call for
a leader election  to name a new leader. By design, 
the leader has a more demanding workload: it pushes writes
to all available  followers, tracks which
followers have confirmed the most recent writes, and
dispatches reads to these confirmed followers.
The leader is the performance bottleneck of Raft.

When workload demands overwhelm the leader, the
Raft infrastructure must acquire and use new
resources to improve its throughput.
One approach is to run the leader on more
powerful computers (\emph{scale up}).
When more powerful computers are unavailable, the leader's load must be split
(\emph{scale out}).
Herein lies the problem: \emph{Raft permits one leader only}.
One leader is essential to make the Raft algorithm
understandable, correct, and easy to implement.
Multi-Raft~\cite{howard2015raft}  scales out by replicating leaders and
followers, and splitting data between
replicas.  Each replica implements a Raft,
providing strong consistency.  Between replicas,
a 2-phase commit provides strong consistency.
Multi-Raft can improve throughput as workload demand
increases, but it is undesirably expensive.  Each scale out
operation doubles resource footprint.

With the global proliferation of networked mobile devices,
follower nodes in Raft are increasingly geo-distributed.
Consider global, coordinated release of some video
content,  Raft coordinates when such content is
accessible. Placing followers in
geo-distributed data centers ensures low  access latency
  for all users.  On the other hand, the cost and amount of cloud
resources requested by followers varies from site
to site.  By na\"ively
replicating leaders and followers, the Multi-Raft
approach inevitably scales out at expensive sites.  Our
research seeks a solution that selectively excludes
expensive sites during scale out, without
compromising throughput or latency.

Multi-leader Paxos algorithms allow multiple
leader threads to process writes.
These algorithms do not suffer follower-leader
workload imbalance like Raft.
However, these algorithms require complicated
procedures to handle  partitions, add
 threads, and remove threads.
Subtle programming mistakes can invalidate strong
consistency guarantee, leading to costly bugs.
Raft allows a single leader only by design, simplifying
programming for Raft.  Such one leader policy
is a key feature of Raft that must be preserved.

In all, there are three main challenges in designing a practical Raft algorithm for large-scale platforms, namely \emph{strong consistency} (i.e., the basics of a Raft algorithm), \emph{cost efficiency} (i.e., avoid temporally expensive sites), \emph{simplicity} (i.e., one leader scheme). To address these challenges,
we propose Black-Water Raft, or \emph{BW-Raft}, a Raft extension that
scales out well with geo-distributed resources. For strong consistency at scale out,
BW-Raft supports 5 types of software threads:
followers, candidates, leader, secretaries and observers.
Inherited from the original Raft, BW-Raft permits zero or one
leader based on majority votes by followers (\emph{simplicity}).
In BW-Raft, leader outsources a portion of its
routines to secretaries, reducing workload imbalance.
Observers answer read-only queries, reducing
workload for followers.   BW-Raft allows multiple
secretaries and observers to run simultaneously.
BW-Raft preserves the safety guarantee of Raft.
We employ \emph{state irrelevancy}~\cite{jong2005state} to prove that
secretary and
observer failures
do not affect correctness.

In BW-Raft,
secretaries and observers are stateless
and execute at any geo-distributed site.
They can be deployed incrementally to achieve high
throughput at a low cost as demanding workload grows.
We present an online approach to discover global, high throughput, and low cost configurations for BW-Raft.
This online algorithm allows BW-Raft to ``peek and peak'', as rafting in a blackwater.  BW-Raft always puts a default node configuration for the arrival workload (i.e., peek), and then reconfigures it after a fixed period, for cost optimization (i.e., peak).
In this optimization, 
our approach accepts time sensitive
parameters on instance expense and workload latency.
This allows BW-Raft to exploit cheap geo-distributed
resources on spot markets, i.e., cloud markets with
cheap but failure prone resources.
For example, in Amazon AWS EC2, spot instances can cost
90\% less than on-demand
instances~\cite{xu2018blending,niu2017handling,niu2015hybrid, yi2016flexible}. 
Though with  cheaper price, introducing unstable 
spot instances into  Raft infrastructure 
can hurt overall reliability. To avoid fatal failures due to spot instances
while harvesting cost benefits,
BW-Raft uses safe, on-demand
instances for leader and followers, and spot
instances for temporary secretaries and observers only.
In the geo-distributed setting, BW-Raft is cheap \emph{per se}, as it leases
spot instances from the cheapest %available
market combined (i.e., cost effective).

We deployed BW-Raft on Amazon EC2
 for more than 12 months serving key-value lookups based on Google and Alibaba
traces~\cite{gcluster,aliTraces}.
BW-Raft purchases low-cost spot
instances in geo-distributed sites to scale out
the performance within the budget.
BW-Raft adaptively boosts read and write
throughput which incurs less than 85\% overhead
compared to Raft~\cite{ongaro2014search}. BW-Raft
scales in increments 5-7X
smaller than Multi-Raft, the state of the
art.
%Geo-Raft
%scales in increments 5-7X smaller than
%state-of-the-art Raft implementations.
BW-Raft reduces costs by 84.5\%, as compared to
Multi-Raft and improves goodput (i.e., the application-level throughput) of $95^{th}$ percentile SLO by 9X. %, compared to state-of-the-art non-adaptive Raft  approaches.
%To the best of our knowledge, we are one of the first few paper implementing strong consensus Raft  scaling out in failure-prone resources.
BW-Raft operates key-value services for 12 months without losing data or crash.

\noindent{This paper contributes as follows:}
\begin{itemize}[leftmargin=0.1in]
\item[-] We identify the scale out performance problem of Raft, a widely used  consistency algorithm, and how today's solution fails at expensive costs. 
\item[-] We propose BW-Raft, an extended Raft design that achieves cost efficiency by exploiting cheap but unreliable geo-diverse spot instances.
\item[-] We prove  in theory that
BW-Raft achieves strong consistency between nodes, and preserves the single-leader policy of Raft.
\item[-] We present an online approach to
  manage  the BW-Raft infrastructure.
\item[-] We built BW-Raft and deployed it for 12 months on public clouds, reporting that
  BW-Raft reduces costs by 84.5\% and improves goodput by 9X, compared to state of the art mechanisms.
\end{itemize}

The remainder of this paper is organized as
follows.
\SEC{SEC:BAC} overviews Raft and the Cloud market.
\SEC{SEC:DES} describes  BW-Raft and
proves that it retains safety guarantees of Raft.
\SEC{SEC:IMP} describes BW-Raft implementation  over geo-distributed spot markets.
\SEC{SEC:EVA} illustrates the BW-Raft can
greatly reduce the cost.
\SEC{SEC:REL} provides a brief on
consistency algorithms and related work.
At last, \SEC{SEC:CON} concludes the paper.

% missing a summary paragraph on connecting the dots of questions and proposes the solution.
\begin{figure*}[t]
  \centering
  \includegraphics[width=6.2in]{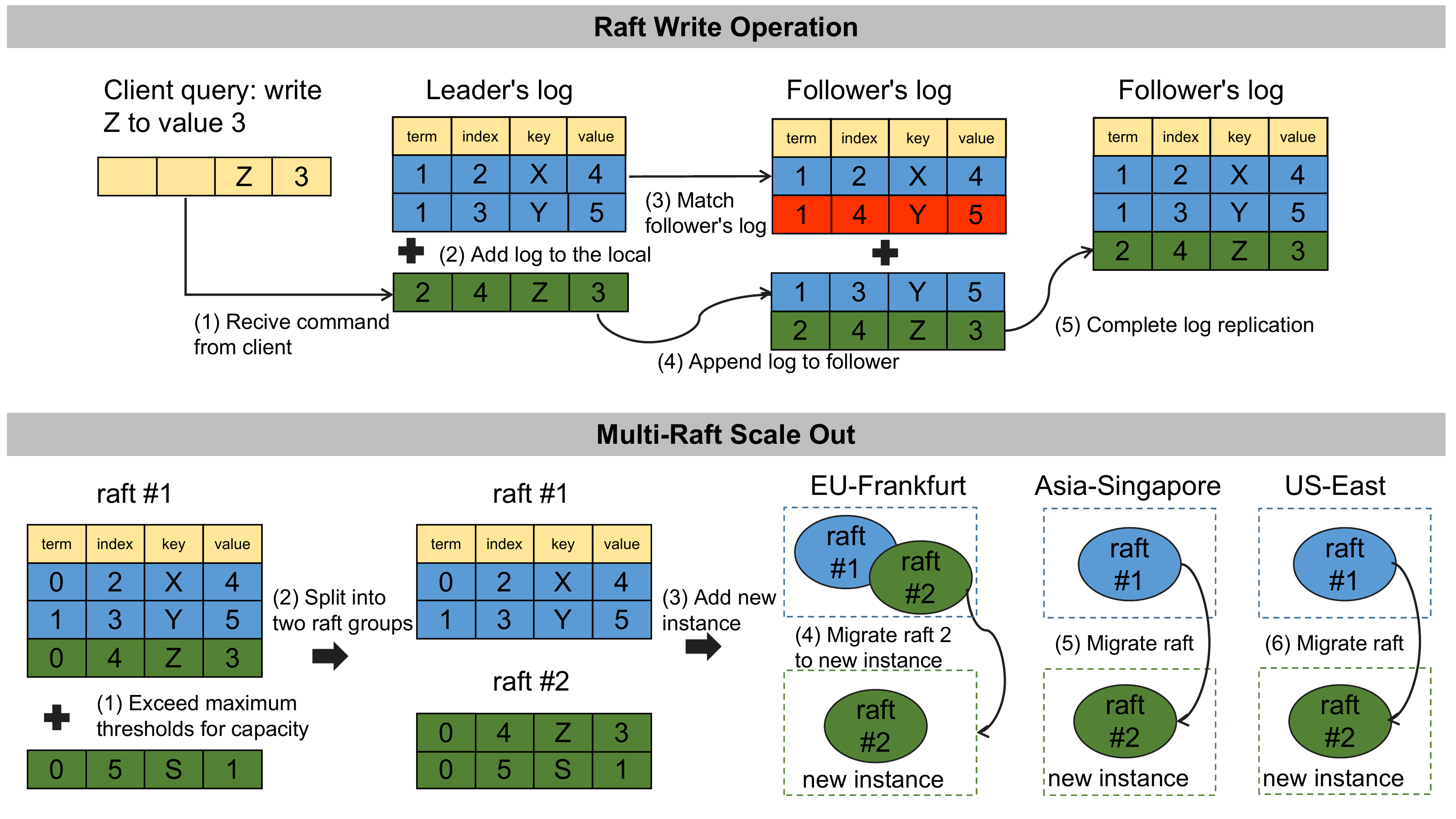}
  \caption{Execution of writes in Raft (top) and
    geo-distributed scale out in Multi-Raft
    (bottom).}
   \label{f:raft-draw}
\end{figure*}

\section{Background}\label{SEC:BAC}

\subsection{A Consensus Algorithm}
In the past decade, Paxos\cite{lamport2001paxos} always was the go-to algorithm to build distributed storage service until Raft~\cite{howard2015raft} was proposed. Due to 
the difficulty of understanding a  Paxos implementation, it is not practical to build a complete Paxos system that is clean and easy to understand from a third party verification. Researchers promote Raft  as it can ensure the linearizability as in Paxos and its notably simplicity and easy to verify~\cite{ongaro2014search}.  Raft has attracted attentions from open source communities in data management, sucessful projects include ETCD~\cite{etcd} and TiDB~\cite{tidb}.

Raft is usually used to implement a key-value store for data intensive services, as follows:
\begin{itemize}
\item {$\mathrm{revision~id \leftarrow write(key~k,value~v)}$}
\item {$\mathrm{\{value~v,~revision~id\} \leftarrow read(key~k)}$}
\end{itemize}

\noindent Raft is designed to handle all  queries  by the leader node, which sets a global processing order for read/write queries and ensuring that subsequent
 queries could return the same value.
In other words, Raft supports linearizable
consistency.

In Raft, we assume all nodes can fail inevitably, like hardware failures, some node's logs can diverge from other's. 
Raft uses a leader-follower structure to
ensure that all nodes always agree on the  leader.
As Figure~\ref{f:raft-draw} (top) shows, there is a conflict between
followers' and leader's log. The leader heartbeats the ``victim'' followers, keeping checking the log conflict and overwriting them until all logs are in consistent.
Besides, Raft forces followers to accept log append from the leader. For example, some clients send  a request to the leader. The leader saves the message locally and broadcasts it to all followers. 
The leader only responds to the client when  the majority nodes have successfully replicated the log.

Raft maintains its consistency against failure through an election process.
In Raft, software threads are either the leader,
followers, or candidates.  The leader handles all the requests from clients and maintains its role by sending heartbeat messages. After receiving a heartbeat, followers reset a random election time. If a follower don't receive a heartbeat after the election time, the follower will increment its token, announcing its candidate role, and call for a election.  Other followers can only vote for a candidate whose log token is not larger than (greater or equal) their own. If a candidate gets the majority votes, it is elected to be the new leader.

With this design, the 
Raft algorithm still remains room to
improvement. There are many researches on  optimizing Raft algorithm. For example, Craft~\cite{wang2020craft} was introduced to to reduce network cost by erasure coding\cite{chen2017giza} . In this paper, we target at the scale out performance of a Raft with concern on cost efficiency. To the best of our knowledge, we are the first work on introducing Raft with the cloud spot  market.

\textbf{Scaling out in Raft.}\label{sub:multiraft}
Raft permits only one leader at a time and the leader has a demanding workload.  Na\"{i}vely
adding follower nodes does not improve throughput.  Instead, it often degrades throughput
by causing more log matching~\cite{xu2019elastic}. Multi-Raft is a widely used approach to scale out Raft.
As shown in Figure~\ref{f:raft-draw} (bottom), it sets up multiple Raft services, splits the key
space and assigns each split to one Raft. Multi-Raft then migrates each Raft service to its
own resources. In geo-distributed cloud settings, each Raft service uses instances at each site.  For example, Figure~\ref{f:raft-draw} shows migration to new instances at AWS EU-Frankfurt, Asia-Singapore, and US-East sites. Each Raft is independent and handles their own update, log commit, and leader elect tasks (e.g., Raft \#1 and Raft \#2 in Figure~\ref{f:raft-draw}). Between leaders of these Rafts, they communicate and keep in consistency based on the 2-Phase Commit. The Multi-Raft approach preserves linearizability within key ranges and scales out.  However, it is not cost efficient.  Scale-out can double the resource footprint to resolve light bottlenecks at only one leader node~\cite{xu2019elastic}.

\subsection{The Cloud Spot Market}
In today's cloud market, spot instances~\cite{awsec2} are short-lived instances offered by cloud providers for a very low cost compared to on-demand or reserved instances. Since customers' demand for cloud resources is dynamic, cloud providers use spot market  to provides excess resources to customers to monetize their excess capacity. The price of spot instances vary with the supply and demand, but on average, users can save up to 90\% compared to on-demand instances. Although spot instance can save lot of money for users, the instance can be interrupted at any time when it's price exceeds the maximum price the user willing to pay. With the growth of cloud services in recent years, more and more cloud providers have launched their own spot instance services to maximize resource utilization and revenue, such as IBM's Transient Servers, Azure's Low-priority VM and Google Cloud's Preemptible VM. In the past, however, spot instances can only provide unstable services, which makes the use of the spot instance very limited. In our paper, we use spot instances to combine the original Raft protocol and extend it into a protocol that can use unstable instances to reach stable services.

\section{Design} \label{SEC:DES}
In this section, we describe the overall design
of BW-Raft. As aforementioned in~\SEC{sec:INT},
\Sysname handles the Raft
protocol across a list of sites to
provide data services with strong consistency.
Such performance is obtained using our
\Sysname mechanism. \Sysname operates
original Raft among sites while hiring spot instances as secretaries and observers to offload
jobs from the leader and followers, respectively.
As such, we allow one leader to handle
a much larger number of followers, as compared to the original Raft.

\begin{figure}[t]
	\centering
	\includegraphics[width=3.2in]{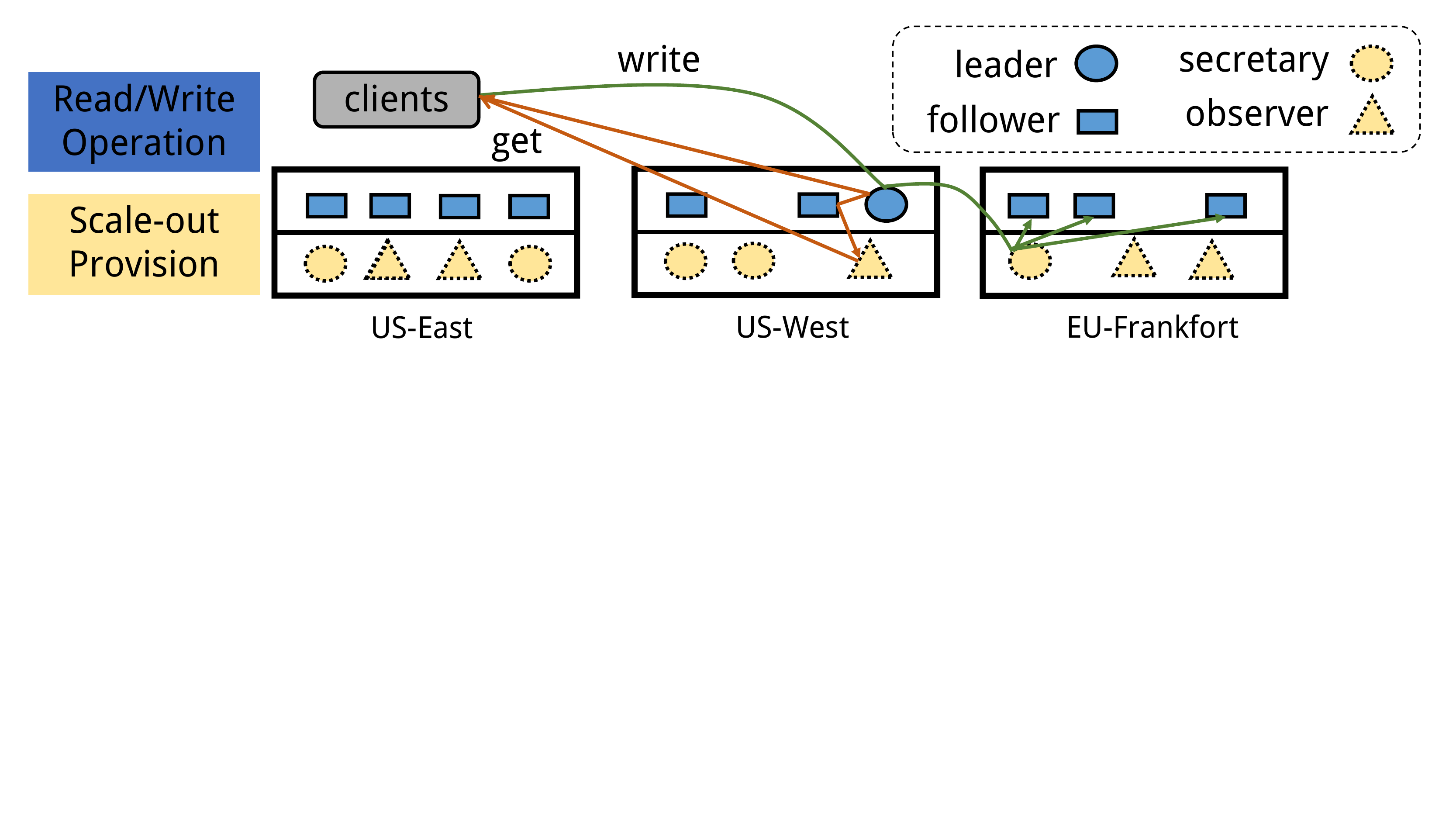}
	\vspace{-1in}\caption{BW-Raft scales out on spot market resources.}
	\label{f:archi}
\end{figure}

BW-Raft extends Raft with two new types of
software threads: \emph{secretary} and \emph{observer}.
Secretaries offload the heavy log appending and log checking jobs from
the leader. Observers relieve read
requests from followers.  Secretaries and
observers are stateless, allowing for elastically scale up and down.
BW-Raft ensures linearizable consistency.  Like
Raft, BW-Raft allows only one leader thread.
The leader is selected from followers as the original election. Tentative nodes, secretaries and observers, can not vote. As such, BW-Raft supports any number of secretaries and observers. 

Secretaries and observers can run on cheap,
failure-prone resources, making them well
suited for
geographically distributed spot markets, i.e.,
cloud resources that are heavily
discounted but can be revoked at any time.
Figure~\ref{f:archi} provides an overview of our
BW-Raft.  Clients issue write queries to the
leader.  The leader offloads log matching and
related tasks to a secretary which runs on a spot
instance.  Read queries are handled by either
followers or observers.  Note, the number of
secretaries and observers varies from site to site
and fluctuates. % as spot prices change.
This gives our design the feasibility of chasing after cheap spot markets.
%As such, we allow higher
%scalability with a much cheaper expense.

The rest of this section first details  the algorithm on the leader election, secretary and observer failures, and the main safety guarantee.
Then, we present a modeling approach for managing the global cloud resources effectively.

\subsection{The BW-Raft Algorithm}
\begin{figure}[t]
\begin{center}
	\includegraphics[width=3.3in]{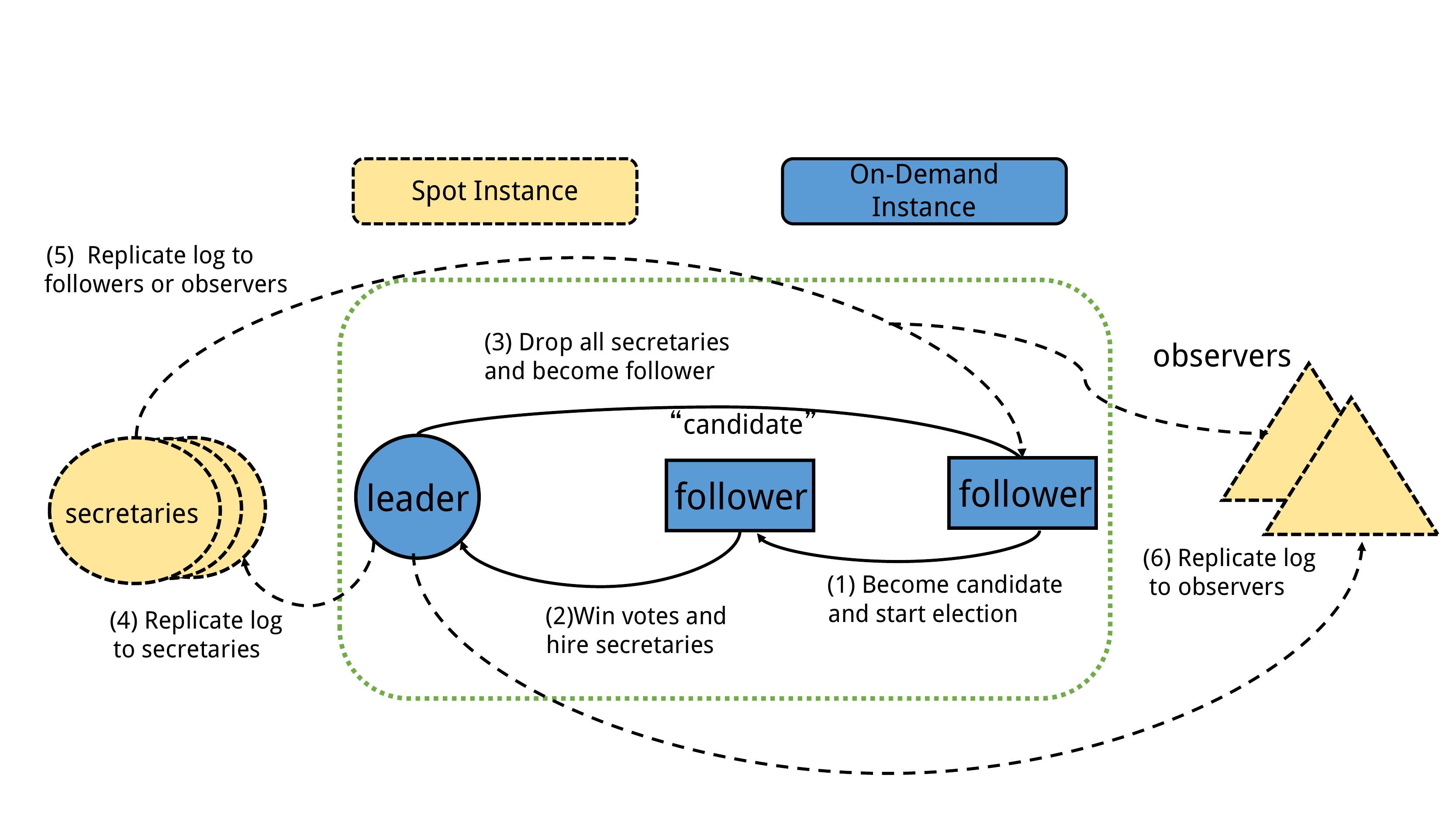}
	\caption{The state machine and stateless nodes in  BW-Raft. The central part in rectangle preserves the classic Raft state machine.}\label{f:state}
\end{center}
\end{figure}

BW-Raft, like the original Raft, initializes cloud
instances at all sites and runs leader and
follower software threads on them.
Figure~\ref{f:state} shows the whole state transfer in
BW-Raft, starting from leader election.

\noindent\textbf{BW-Raft Leader Election}\label{sub:elect}:
The leader manages log matching and replication
for followers and secretaries.
The execution trace of commands for the state
machine is based on logging.\vspace{5pt}

\noindent PROPERTY 3.1. \textbf{(Leader Election Safety).}  \textit{Like Raft,  there is at most one leader per term in BW-Raft.}

The leader maintains its role by sending heartbeat messages. After receiving a heartbeat, followers set a random timer.  If a follower does not receive a message
before the timer triggers, the follower calls for
an leader election and stops all secretary threads  (i.e., Step (1) in Figure~\ref{f:state}).  The follower increments it's term and tells other followers that it is a "candidate" for leader.
Followers vote for a candidate whose log is not older than its own. If election times out, the election will restart again (not shown in Figure~\ref{f:state}). If a candidate gets the majority votes from most followers,
this secretary will become a new leader and BW-Raft  provisions secretaries for the new leader
(i.e., Step (2) in Figure~\ref{f:state}).

BW-Raft starts each term ($\bar{T}$) with a leader election. The leader orchestrates normal operations (log  management), notifies higher term when new $\mathrm{update}$, as shown in   Step (3) in Figure~\ref{f:state}. Meanwhile, the leader in BW-Raft always tells followers which secretaries are responsible for this log, such that log management can be offload to assigned secretaries (i.e., Step (5) in Figure~\ref{f:state}).
If the election is fail, a new term starts with a new election. \vspace{5pt}
%%%  Chris commented out below sentence because it
%%%  seems to be
%%%  covered in implementation
%%%
%BW-Raft  operates the   state machine using log
%append/commit with extra operations on
%secretaries, reversing salient features from Raft
%in a geo-diverse environment:

\noindent PROPERTY 3.2. \textbf{(BW-Raft  State Machine Safety)}  \textit{ Each replicated copy of the state machine executes the same commands in the same order.}

Established in Woos and Wilcox's prior research in Raft~\cite{woos2016planning}, this property   provides consistency guarantees for logs between all leader, secretaries, and followers in the protocol.

\noindent \textbf{Stateless Operations in BW-Raft Leader Election}:
Besides the whole leader election process, BW-Raft employs stateless instances as \emph{secretaries} and \emph{observers}. After a successful election,
when the term ($\hat{T}$) is longer than a predefined period ($T < \hat{T}$), leader will re-provision secretaries and offload logs (i.e., Step (4) in Figure~\ref{f:state}) by the end of this period. BW-Raft adopts this re-election mechanism to mitigate the fail-prone feature from spot instances, without incurring noticeable delay/cost.

Followers that operate in BW-Raft can arbitrarily hire cheap local spot instances as  observers. Each time a follower receives a $\mathrm{write}$, it will append this $\mathrm{write}$ to all its linked observers without waiting for the commit. When the follower receives a $\mathrm{read}$, it   reroutes the job to committed observers. If no observers commit previous appending log, this follower
  executes the $\mathrm{read}$ itself (i.e., Step (6) in Figure~\ref{f:state}).

\begin{figure}[t]
\begin{center}
	\includegraphics[width=3.3in]{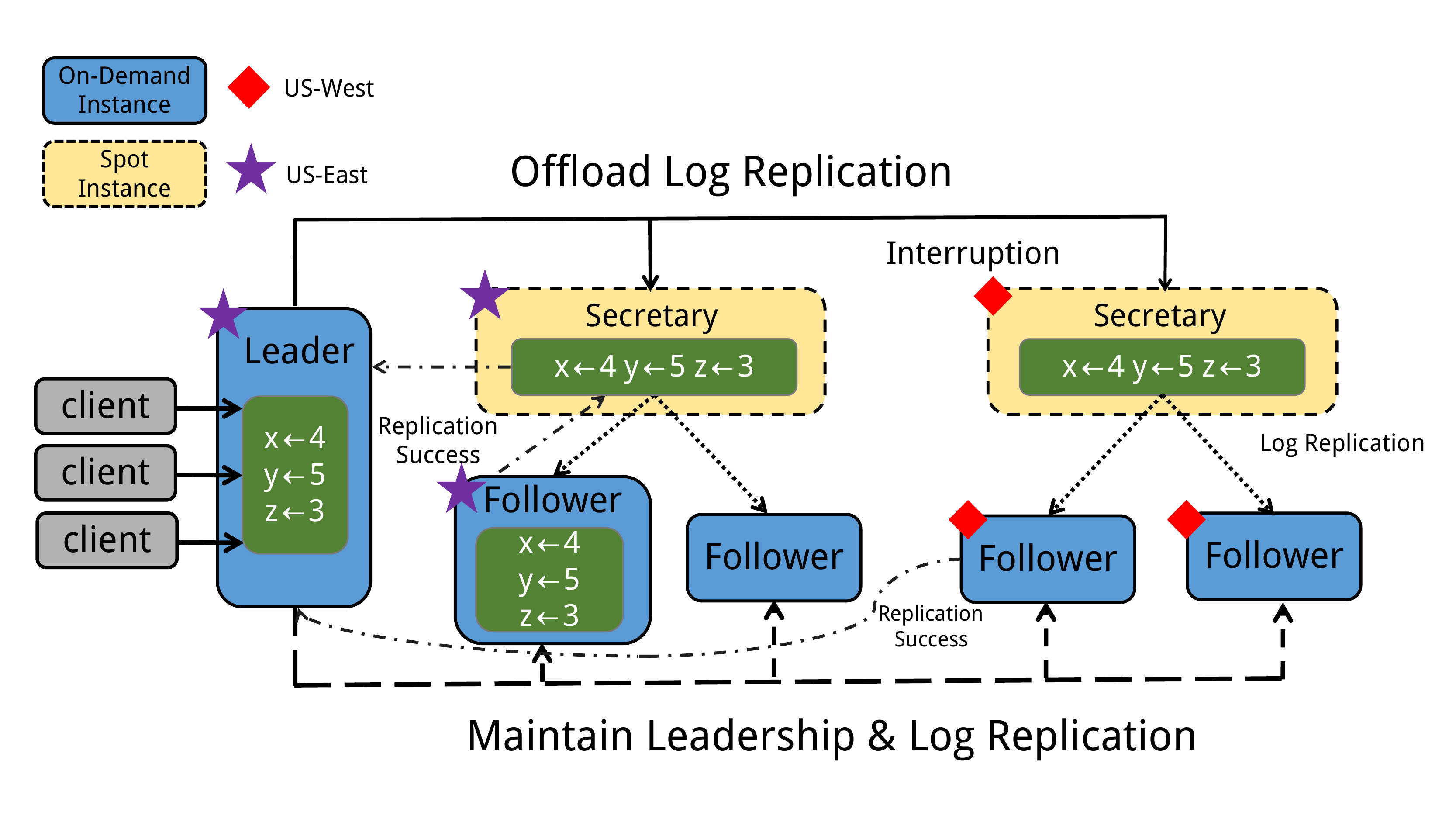}
	\caption{Secretary offload log replication.}\label{f:sec}
\end{center}
\end{figure}

\noindent \textbf{Log Replication:}
BW-Raft targets at optimizing the performance in the log replication phase.
In log replication, the list of commands to execute on the state machine is kept in log, and the position of a command in the log is called its index. Each node has its own copy of the log from leader, and state machine safety reduces to maintaining agreement between all copies of the log.
Figure~\ref{f:sec} shows an example of BW-Raft. When a client sends a $\mathrm{write}$  to the leader, the leader first appends a new log entry containing that command to its local log. Then the leader sends an appended message containing the entry to an assigned secretary, which is responsible to replicate the log to the followers in the same site. Each follower in the same site appends the entries to its log and sends the acknowledgment to the assigned secretary. To ensure that followers' logs stay in consistent with the log of the leader, messages include the index and term of the previous entry in secretary's log; the follower checks whether it has an entry with the index and term before appending new entries to its log. This consistency check guarantees the following property: \vspace{5pt}

\noindent PROPERTY 3.3. \textbf{(Log Matching).}  \textit{ If any two logs contain entries at a particular index and term at different geographical locations, then the logs are identical in BW-Raft.}

 After the secretary do a log replication operation, it will report the number of follower that have agreed, and the leader commits the write based on the cumulative amount of agreed followers. Once the leader learns the majority followers have acknowledged the new entry, it will executes the command contained in the committed entry on the state machine and responds to the client with the output of the command. The followers and observers are informed by the leader's heartbeat that they can safely execute the command on their state machines. Thus, in both leader election and log replication, both secretaries and observers are state irrelevant.\vspace{5pt}

\noindent PROPERTY 3.4. \textbf{(State Irrelevancy).}  \textit{ If any secretaries/observers fail during the term at different locations, then the logs are still matched because state machines in both leader and followers are irrelevant to secretaries and observers.}

We first prove the State Irrelevancy in secretaries. Given an execution trace $\tau$ of BW-Raft, assuming one secretary fails during the trace,
we shall find an existing $\sigma$ such that $\tau$ linearizes to $\sigma$. To find the $\sigma$,
we revert BW-Raft back to a single Raft, then we   pick the sequence of commands executed by the followers on their local state machines. State machine safety guarantees that all nodes agree on this sequence.

The remaining job is to show that without failing certain secretaries,   we have $\tau'$
 linearizes to $\sigma$. Let $\tau'$
be the sequential input–output trace corresponding to $\sigma$ in $\tau$. That is,
for each command executed in $\sigma$, $\tau'$
contains an input immediately followed by
the corresponding output for that command. All commands from/to secretaries can be omitted in
$\tau'$, which makes it a permutation of $\tau$  that respects the ordering condition
of properties. Each of these is established as a separate invariant
by induction on the execution.
The proof for observers are similar since they never affect  the order of local state machine of the correspondent followers.

With all properties inherited from Raft, while the additional roles in BW-Raft  are state-irrelevant, we have, \vspace{5pt}

\noindent THEOREM 3.5.\textbf{}  \textit{ Properties 3.1–3.4 imply linearizability in BW-Raft.}

The proof trivially follows the linearizability proof in Raft~\cite{woos2016planning}.
Once an entry is committed, it becomes durable, in the sense that
its effect will never be forgotten by BW-Raft. In
all, BW-Raft  operates log replication with
linearizability.\vspace{5pt}

\noindent PROPERTY 3.6. \textbf{(Linearizability).}  \textit{ BW-Raft implements a linearizable state machine.}

The proof trivially follows the linearizability proof in Raft~\cite{woos2016planning}.
BW-Raft  also provides a liveness guarantee: if there are sufficiently
few failures, then the system will eventually process and respond
to all client commands.

\subsection{Global Resource Management}\label{sub:modeling}

BW-Raft offloads  on secretaries and observers, by exploiting cheap but revocable resources such as spot instances for cost efficiency. This section answers a critical infrastructure question, for a given data service, how many and which sites the required  instances are purchased. This is the global resource management in BW-Raft.

BW-Raft infrastructure manages its global resource via ``\emph{peek and peak}''. That is, whenever a new data service arrives, there is a stationary configuration for it, i.e., the original Raft. In other words, BW-Raft runs under the original Raft model at the beginning of new service requests. Once the target service load increases beyond the default capacity, BW-Raft starts to scale, predict, and assign new resources, as rafting in black water,  or ``\emph{peek}''. For example, BW-Raft runs a monitor process to ``\emph{peek}'' the resource demand, instance statistics, and workload profiles. On the other hand, if the target service scales out, we would like to pick at least $k$ nodes to meet the demand, and select the top-$k$ best cost-efficient instances online. This decision recurs every epoch, which allows   services ``\emph{peak}'' elastically.

%the peek process, or the optimization process
% decide k static
\noindent \textbf{Peek in BW-Raft}.
Given a newly arrived service, BW-Raft configures and runs a stationary setup for a fixed period of time $T$, e.g., 1 hour. The value of $T$ should be set according to the SLA requirement. During this period, we collect request counts, types, cluster dynamics, and prices of all sites in the past $T$ time. Based on this information, BW-Raft can calculate the number of  spot instances needed for expansion by algorithm~\ref{alg::calculateK}.

BW-Raft collects statistics every period time $T$ and dynamically manages  global resources while the system is running. Supposing one secretary can manage $f$ followers, then we can get the total number of secretaries needed to offload leader's workload in current cluster($k_s'$, line 3) and the number of new secretaries($\Delta k_s$, line 4). Note that if there are $F_i$ ($\frac{f+1}{2} \le F_i \textless f$) in a data center, actually that data center should also need one secretary. BW-Raft prioritizes the number of secretaries or observers based on the read-write ratio of workload. For example, if the write ratio $\zeta$ in current period time  is less than $\varpi$=30\%, BW-Raft  gives the priority to the number of observers(line 5-15).  Otherwise, the number of secretaries is given priority(line 16-20). In fact, $\varpi$ is user-defined variable in practice.

When prioritize observer, we need calculate the growth rate of read requests $\mathcal{A}$(line 6) firstly. That's because observers are only used to processing read requests. As a result, the number of observers is closely related to the amount of read requests. Taking fluctuations of systems in the real environment into account, we argue that it is unnecessary to rent new observers or withdraw old observers when $|\mathcal{A}| \le 10\%$, for the sake of avoiding extra overhead of frequently schedule resources. If $|\mathcal{A}| \textgreater 10\%$, it means that we need more observers to address increasing read requests. Basically, we provide one observer for each data center. However, subject to the constraint of budget $\vartheta$ and unit price $\rho$, $m$ data centers can hire at most $\Delta k_o$ new observers(line 8-9). If $|\mathcal{A}| \textless -10\%$, it implies that the read requests are reducing, it's not cost-effective to serve such read workload with $k_o$ observers. In that case, we should cut down at most $m$ observers(line 11). Then we use remaining budget to rent secretaries and update available budget(line 13-15).

When prioritize secretary, we first hire at most $\Delta k_s$ new secretaries. Likewise, then hire new observers with the rest budget and calculate available budget(line 17-20).

At the end of algorithm~\ref{alg::calculateK}, it updates the total number of secretaries and observers in cluster(i.e., $k_s$ and $k_o$). Meanwhile, it gets the number of new spot instances needed to scale out for following stage "Peak in BW-Raft".

After that, the estimated total expense can be calculated by equation (1):
\begin{align}
    \label{totalCost}
    cost = \displaystyle \sum_{i=1}^{m}\{\beta F_i\} + \beta + \rho (k_s+k_o) + \mathbb{C}
\end{align}
where $\mathbb{C}$ is a linear function of network cost related to the total number of instances in cluster.

% $\mathcal{A}_r$
\begin{table}[]
    \centering
    \caption{Key Parameters}
    \label{parameters}
    \begin{tabular}{ll}
    \toprule
    Symbol &  Description\\
    \midrule    
        $\rho$  & The unit price of spot instance\\
        $\beta$ & The unit price of on-demand instance\\
        $\vartheta$  & The available budget\\
        $k_s$ & The number of secretaries in cluster \\
        $k_o$ & The number of observers in cluster\\
        $\Delta k_s$ & The number of new secretaries\\
        $\Delta k_o$ & The number of new observers\\
        $k$ & The number of spot instances that need to be rented\\
        $N_r$ & The number of read requests in last period\\
        $N_r'$ & The number of read requests in current period\\
        $\mathcal{A}$ & The growth rate of the number of read requests\\
        $\varpi$ & The write ratio threshold \\
        $\zeta$ & The write ratio in current period\\
        $m$ & The number of data centers\\
        $F_i$ & The number of followers in $i_{th}$ data center\\
        $f$ & The number of followers one secretary can handle\\
        $\mathbb{C}$ & A linear function of network cost \\
    \bottomrule     
    \end{tabular}
\end{table}

\begin{algorithm}[t]
  \caption{The algorithm of calculate k}
  \label{alg::calculateK}
  \begin{algorithmic}[1]
    \Require
      The parameters mentioned in Table ~\ref{parameters}
    \Ensure
      $k$: the number of new spot instance, including secretary and observer 
    \State initial $k_s=0$, $k_o=0$, $\varpi$=30\%;
    \For{every period time $T$}
        \State $k_s' \gets \displaystyle \sum_{i=1}^m  \frac{F_i+ \frac{f+1}{2} } {f}  $
        \State $\Delta k_s \gets k_s'-k_s$
        \If{$\zeta \le \varpi$}
            \State $\mathcal{A} \gets  \frac{N_r'-N_r}{N_r}$
            \If{$\mathcal{A} \textgreater 10\%$}
                \State $\Delta k_o \gets m$
                \State $\Delta k_o \gets min(\Delta k_o, \frac{min(\rho \Delta k_o, \vartheta)}{\rho})$
            \ElsIf{$\mathcal{A} \textless -10\%$} 
                \State $\Delta k_o \gets max(-k_o, -m)$
            \EndIf
            \State $\vartheta \gets max(0,\vartheta-\rho \Delta k_o)$
            \State $\Delta k_s \gets min(\Delta k_s, \frac{\vartheta}{\rho})$ 
            \State $\vartheta \gets max(0,\vartheta-\rho \Delta k_s)$
        \Else 
            \State $\Delta k_s \gets min(\Delta k_s, \frac{\vartheta}{\rho})$
            \State $\vartheta \gets max(0,\vartheta-\rho \Delta k_s)$            
            \State $\Delta k_o \gets min(m, \frac{\vartheta}{\rho})$
            \State $\vartheta \gets max(0,\vartheta-\rho \Delta k_o)$  
        \EndIf
        \State $k_s \gets k_s + \Delta k_s$
        \State $k_o \gets k_o + \Delta k_o$
        \State $k \gets \Delta k_o + \Delta k_s$    
    \EndFor
  \end{algorithmic}
\end{algorithm}

\noindent \textbf{Peak in BW-Raft}.

Spot Instances are quantified using the following weighting formula based on CPU capacity $c$, available memory $\phi$, average price $\varrho$ and spot revocation probability $\xi$, besides,  $\ell_i$ represents their coefficients. The spot revocation probability can be predicted using \emph{RevPred} model~\cite{2020SpotTune}.
{
\begin{align}\label{e:score}
score &=\displaystyle\frac{\ell_1 c+\ell_2 \phi+\ell_3\frac{1}{\varrho}}{\xi}
\end{align}
}

Through data statistics and calculation in the ``peek'' stage, we know that we need to rent $k$ spot instances. Now, each spot instance has a score in spot instance pool. Using MCSA(\emph{Multiple-Choice Secretary Algorithm}) algorithm~\cite{topk-xu}~\cite{topk-original}, we can select top-$k$ spot instances. MCSA is an online algorithm that picks top-$k$ elements and maximize their sum in a dynamically varying set of elements after checking a preliminary list of observations, i.e., peek. The complexity of this algorithm on picking the top-k is O(n).  

\begin{algorithm}[h]
  \caption{MCSA:get top-k spot instances}
  \label{alg::getTopK}
  \begin{algorithmic}[1]
    \Require
      $score[]$: array of spot instances's score;
      $k$: the number of spot instances selected;
    \Ensure
      $topk[]$: array of selected spot instances's index
    \State initial $L=1$ and $R=len(score)$;
    \Function {TOP-K}{$k, L, R, score$}
        \If {$k>1$}
            \State $m \gets binornd(R-L+1,\displaystyle \frac{1}{2})$;
            \State TOP-K($floor(\displaystyle \frac{k}{2}),L,L+m-1,socre$);
            \State TOP-K($k-floor(\displaystyle \frac{k}{2}),L+m,R,socre$);
        \Else
            \State $len \gets R-L+1$;
            \State $n \gets floor(\displaystyle \frac{len}{exp(1)})$;
            \State $mx \gets score[L]$;
            \For{$i=L \to L+n-1$}
                \State $val \gets score[i]$;
                \If {$val>mx$}
                    \State $mx \gets val$;
                \EndIf
            \EndFor
            \For {$i=L+n \to R$}
                \State $val \gets score[i]$;
                \If {$val>mx$}
                    \State $topk.append(val)$;
                    \State \Return{}
                \EndIf
            \EndFor
            \State $topk.append(mx)$;
        \EndIf
    \EndFunction
  \end{algorithmic}
\end{algorithm}

The input of MCSA(algorithm \ref{alg::getTopK}) is an array of spot instance scores and the $k$ mentioned above. The output is an array stored the index of selected $k$ spot instances. If $k=1$, the algorithm finds the maximum number $mx$ of previous floor($n/e$) elements, and continues to search in order until a larger number is found, as compared to $mx$, or if no number larger than $mx$ is found (line 7-25). If $k>1$, the search space is  divided into two groups, find $floor(k/2)$ indices in the first binomial distribution $binornd(R-L+1,1/2)$ elements, and looks up $k-floor(k/2)$ indices in the rest elements. The algorithm repeats until $k=1$ gets one selected index.

For $m$ data centers, secretaries and observers are evenly distributed, according to the number of followers in each data center. For $i_{th}$ data center, assign at most $\frac{F_i+\frac{f+1}{2}}{f}$ secretaries. When read requests leap, try to add one new observer in every data center. In order to reduce excessive overhead of long-distance communication, when assigning secretaries and observers, give priority to nodes close to the targeted data center. For areas where the number of followers is less than $f$, the secretary and observer will be allocated when there is a surplus.

For unexpected bursty traffic in cloud, BW-Raft can utilize \emph{PostMan}~\cite{DBLP:conf/usenix/JinGXSNLQW19} to achieve load balance. According to \emph{PostMan}'s architecture, it adds new role \emph{helper} between client and server to assemble small packets into large ones, and then send large packets to heavily-loaded servers. When servers receive assembled packets, they disassemble big packets and running under BW-Raft. The key idea of \emph{PostMan} is motivated by the observation that processing large packets incurs far less CPU overhead than processing small ones.

BW-Raft can effectively achieve fault tolerance due to the state irrelevancy property of spot instances. On one hand, if observer fails, client will time out because it does not get reply within a given amount of time, then client will resend read request to another node. On the other hand, if secretary fails, that secretary can't continuously offload leader's workload. In that case,  workload will return to leader and leader is responsible for rest append log tasks. During the running time of system, BW-Raft will dynamically manage global instances by algorithm \ref{alg::calculateK} and \ref{alg::getTopK}.  In extreme situation(i.e., all spot instances fail), BW-Raft will turn back to Raft. 

Gradually, BW-Raft calculates $k$ based on the statistical information of the previous $T$ time (\emph{peek}), and then the best $k$ spot instances are selected to allocate new secretaries and observers (\emph{peak}). In practice, $T$ is changeable, if $T$ is set to a fine-grained manner, BW-Raft can approach to a real time system if the network bandwith allows.

\begin{lstlisting}[float=tp,caption={Remote  Procedure Call in BW-Raft },captionpos=b]
Roles := Follower|Leader|Secretary|Observer
message Entry {index, key, value}

// Leader|Follower|Candidate 
service BW-RAFT {
    rpc RequestVote(RequestVoteArgs) 
        returns (RequestVoteReply){};
    rpc AppendEntries(AppendEntriesArgs) 
        returns (AppendEntriesReply){};
    rpc GetReadindex(ReadIndexArgs) 
        returns (ReadIndexReply){};
}

// Secretary
service BW-Secretary {
    rpc L2SAppendEntries(AppendEntriesArgs) 
        returns (L2SAppendEntriesReply){};
}
// Observer
service BW-Observer{
    rpc AppendEntries(AppendEntriesArgs) 
        returns (AppendEntriesReply){};
}
// Client|Server
service BW-KV {
    rpc PutAppend(PutAppendArgs) 
        returns(PutAppendReply){};
    rpc Get(GetArgs) 
        returns (GetReply){};
}

\end{lstlisting}

\begin{figure}[t]
\begin{center}
	\includegraphics[width=3.1in]{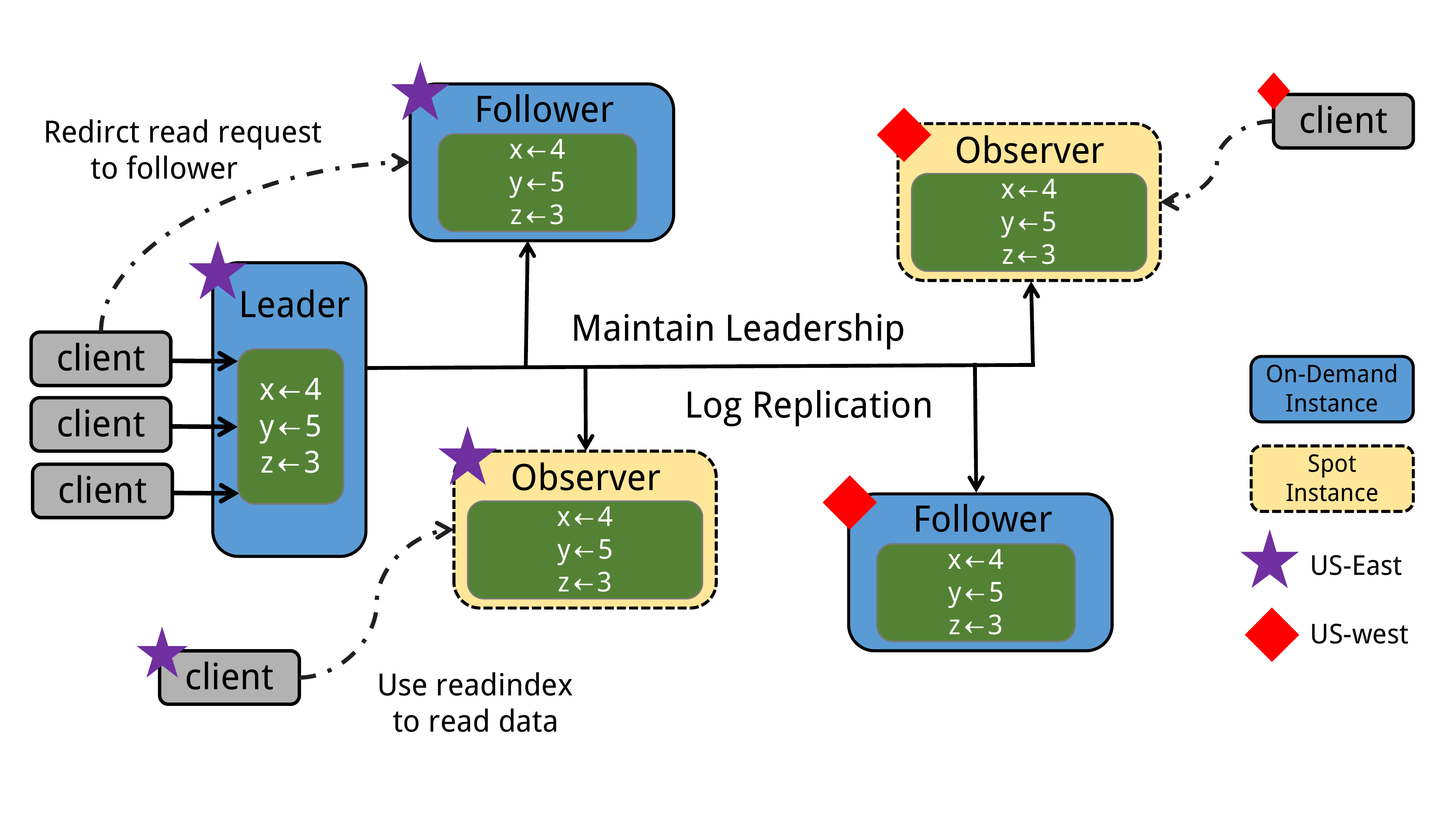}
	\caption{Observer offload log replication.}\label{f:obs}
\end{center}
\end{figure}

\section{Implementation}\label{SEC:IMP}

To evaluate our BW-Raft protocol, we first implement a key-value service based on Raft. Then we modify it to adapt  BW-Raft which  provides several interfaces to enable a scalable geo-diverse replicated state machine. All of the implementation code is more than 4000  lines.  We use leveldb \cite{leveldb} as our storage engine and different node  communicates via grpc \cite{grpc} .

%BW-Raft  realizes several interfaces which enable a scalable geo-diverse replicated state machine.   First, BW-Raft  provides an interface that implements strong consensus  between  sites. Among all sites, BW-Raft  leases global and local spot instances as secretaries for leaders. Within a site, BW-Raft hires local spot instances as observers for followers.  The total number of spot instances is dynamically tuned based on current performance statistics and resource provision, as discussed in \SEC{sub:modeling}. BW-Raft  also exposes the management interface  to clients, supporting upper-level applications. Hence, developers could switch to BW-Raft with a few extra lines of codes.

\subsection{Original Raft Implementation}

When BW-Raft doesn't employ secretary and observer, it only includes three roles: leader, follower and candidate, just like the original Raft. So we implement the Raft protocol first and then modified it to allow leader to hire secretaries and observers. As shown in the Listing 1, we defined the  BW-Raft service to provide the functions of RequestVote and AppendEntries. When follower does not receive the leader's heartbeat for a certain period of time, it increases its term and turns into a candidate. Then that candidate starts a new election and use RequestVote RPC to collect votes. If the term and index of the candidate's last log are at least up-to-date as follower's, the follower votes for the candidate, otherwise the candidate's voting request can be rejected. In a round of election, if a candidate collects more than half of cluster members' votes, it becomes a new leader. Otherwise, a new round of election starts. When a new leader is generated, it starts to synchronize the logs with each node and sends heartbeat to prevent new election.
%Recall from Figure~\ref{f:raft-draw}, the data structure used to append entries includes $\mathrm{<term, index, key, value>}$.  

%In order to mamag leader also add a data structure of assigned secretary and  follower. In order evaluate our protocol's performance we then implement kv service to provide read and write operations as  shown in the Listing 1's BW-KV. 

%In our implement of BW-Raft, we defined 4 types of RPC (shown in Listing 1). The first type of RPC allows leaders to start secretaries.  In the second type, the leader uses RPC append follower logs. Recall from Figure~\ref{f:raft-draw}, the data structure used to append entries includes $\mathrm{<term, index, key, value>}$.  The leader  also includes an address for an assigned  secretary in this  RPC. Followers reply to secretaries whom inform the leader when the log has been appended by a majority of followers. As in Raft, leader and follower  RPC piggy backs heartbeat messages to maintain leadership. If there is no secretary in the runtime, leader directly replicates log entries to followers. In this case, BW-Raft reverts and equals a classic \raft. The third and forth RPC type occurs when secretaries offload log management and observers receives reads, respectively.  Secretaries use RPC to discover matching indices within follower logs. Recall, secretaries are designed to run on cloud instances that could fail at any time. We hypothesize that future implementation of this RPC could use a highly frequent asynchronous communication. In our prototype, we used heartbeats  in a short sliding window.

\begin{figure*}[t]
\centering
$\begin{array} {ccc}
\includegraphics[height=0.24\textwidth]{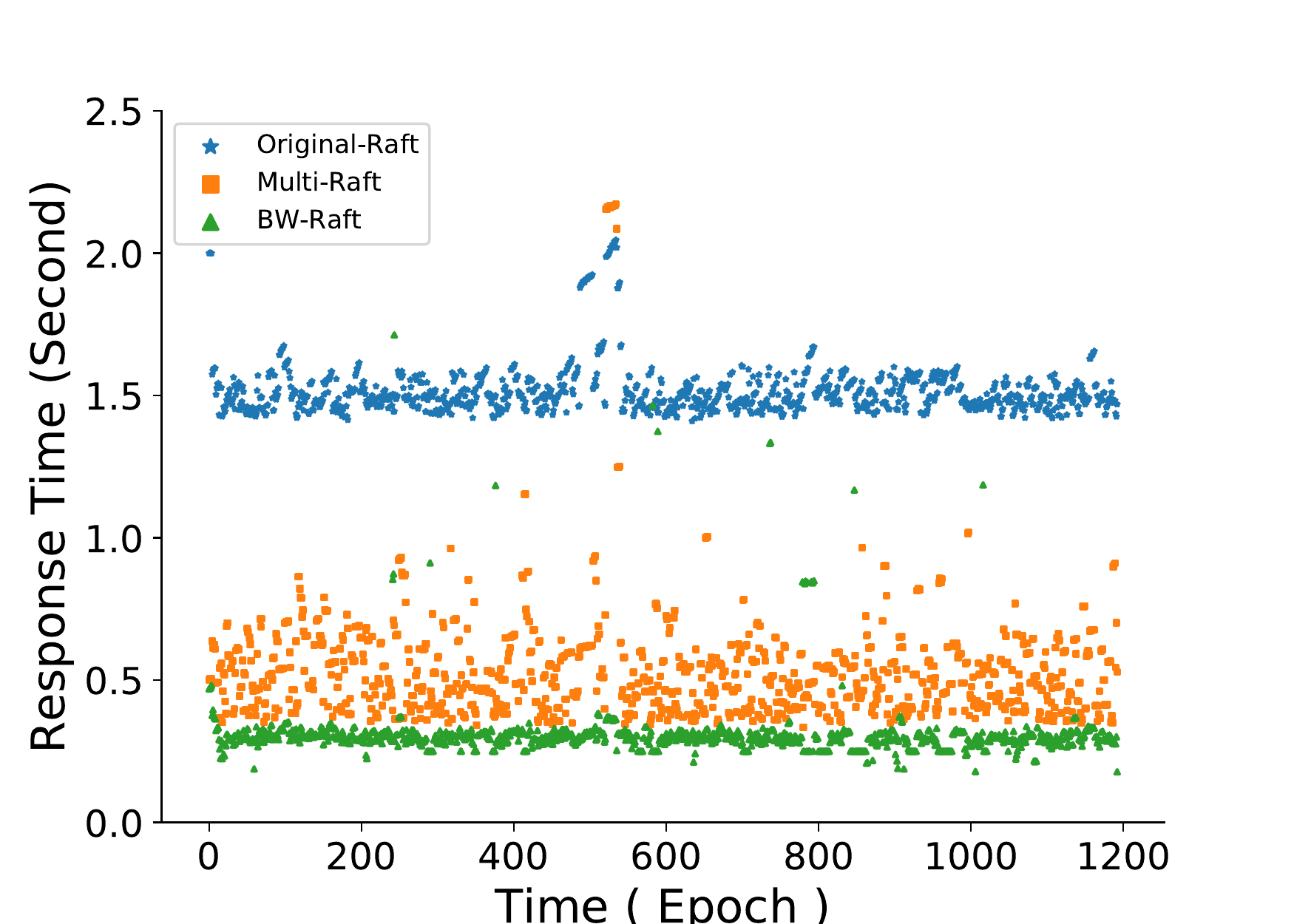} & \includegraphics[height=0.24\textwidth]{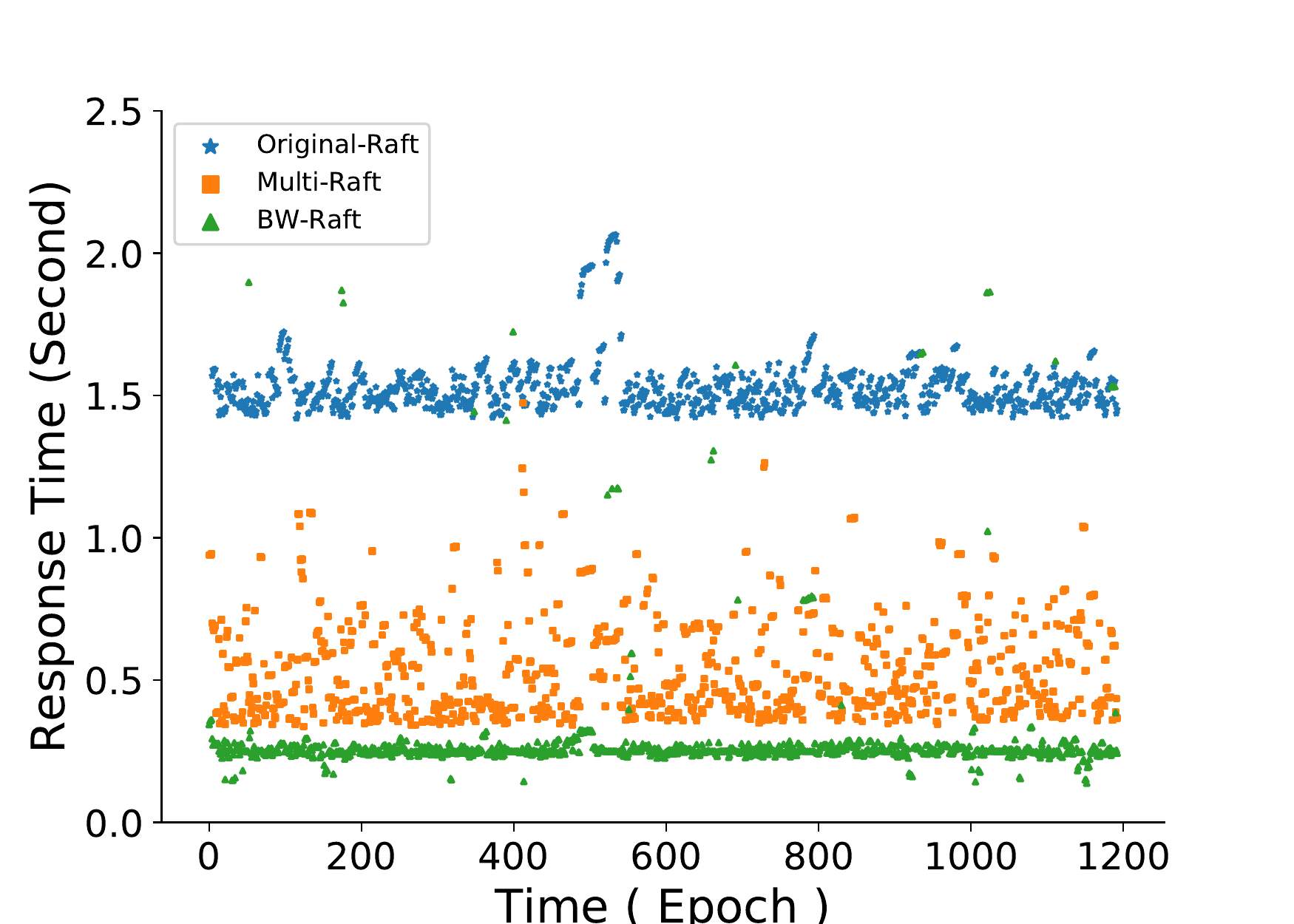} & \includegraphics[height=0.24\textwidth]{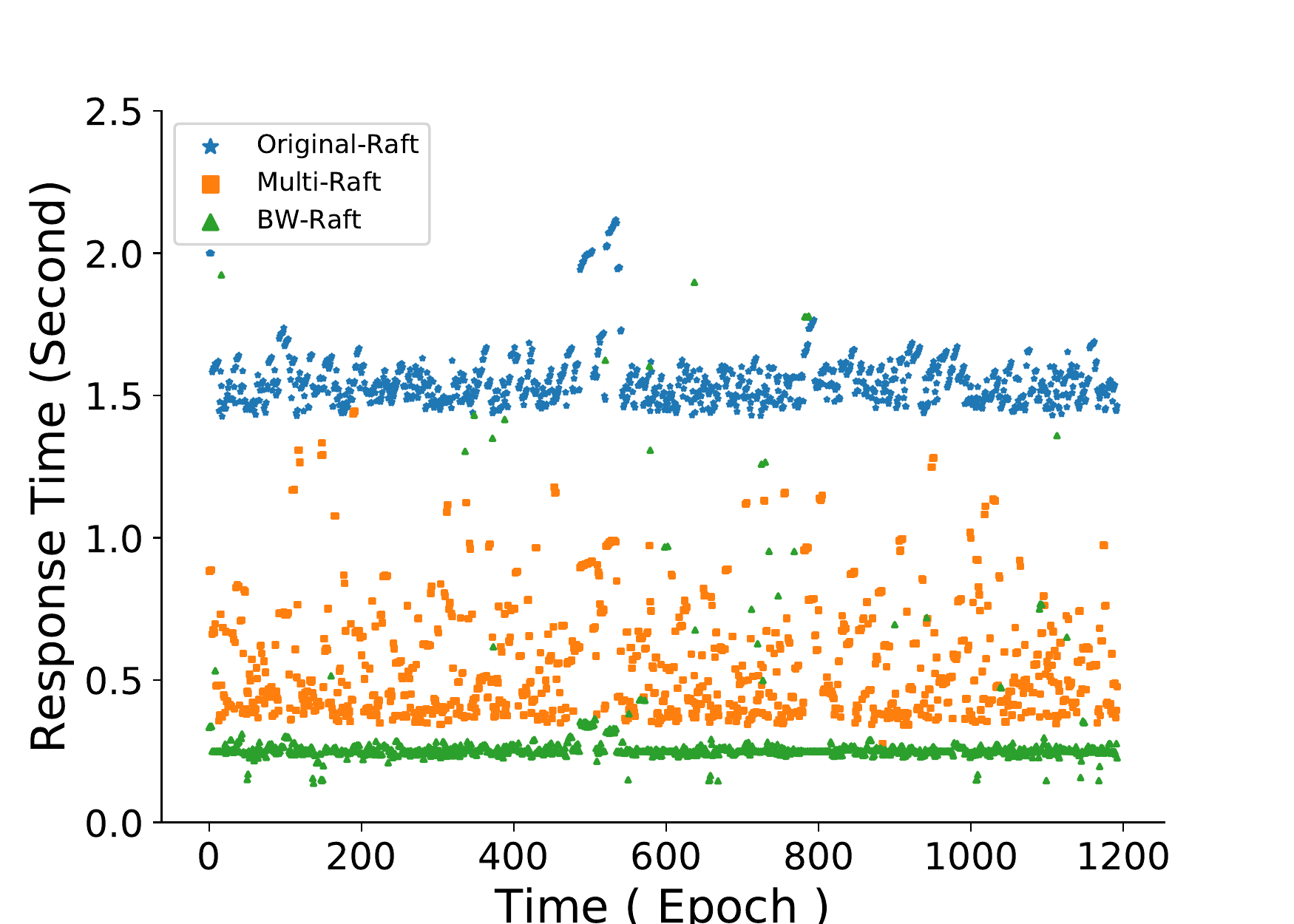} 
\\
\includegraphics[height=0.24\textwidth]{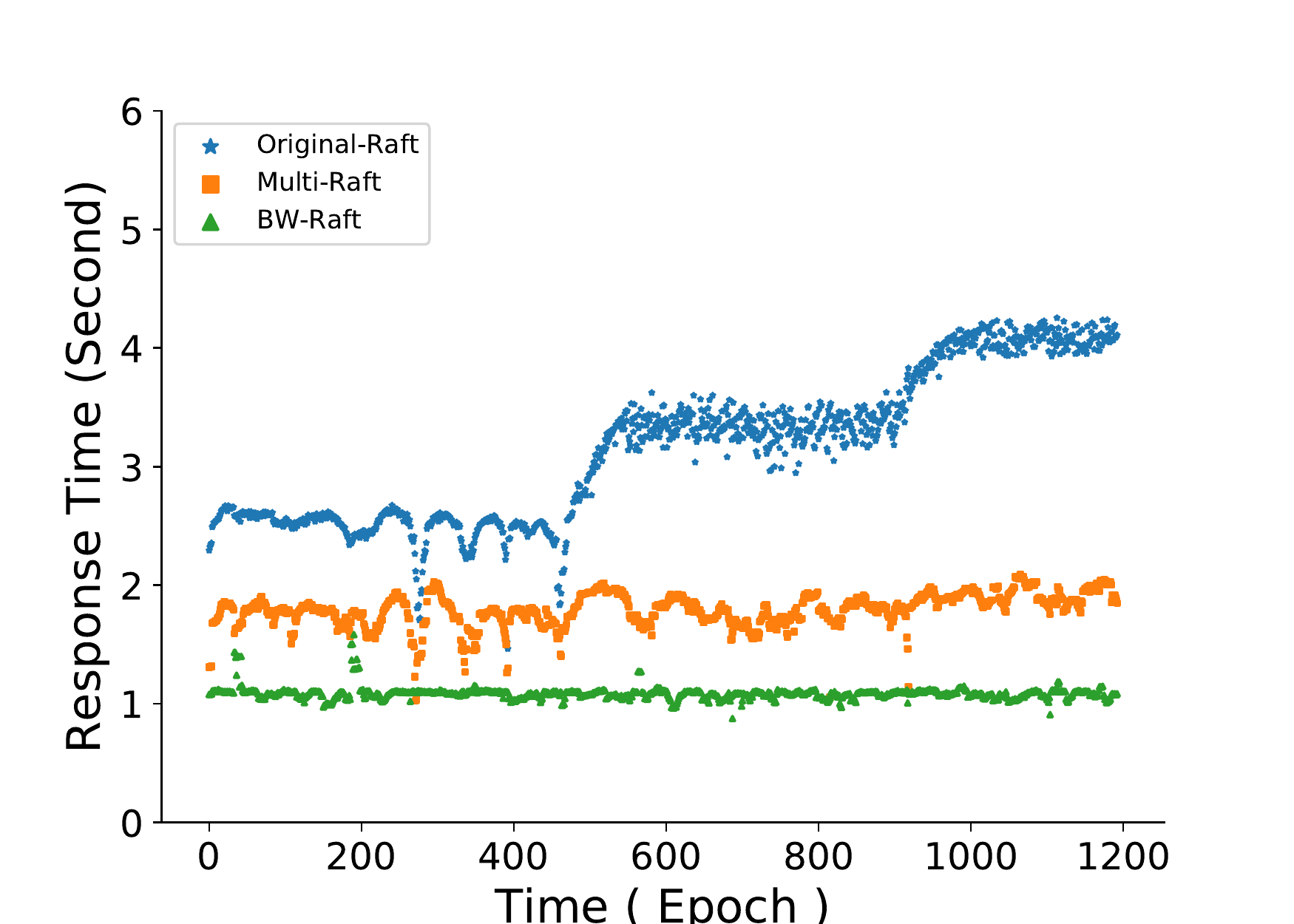} &  \includegraphics[height=0.24\textwidth]{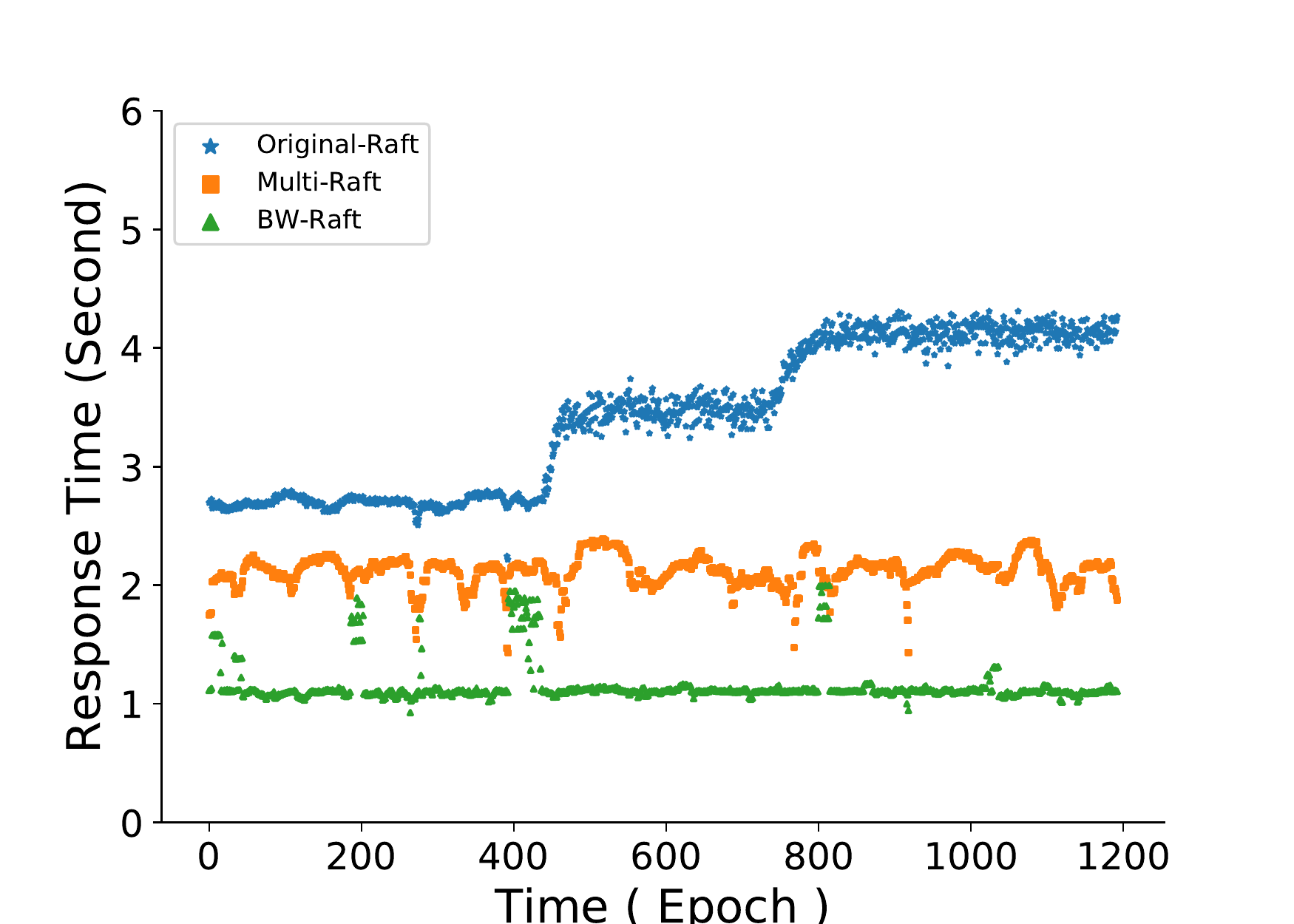} &  \includegraphics[height=0.24\textwidth]{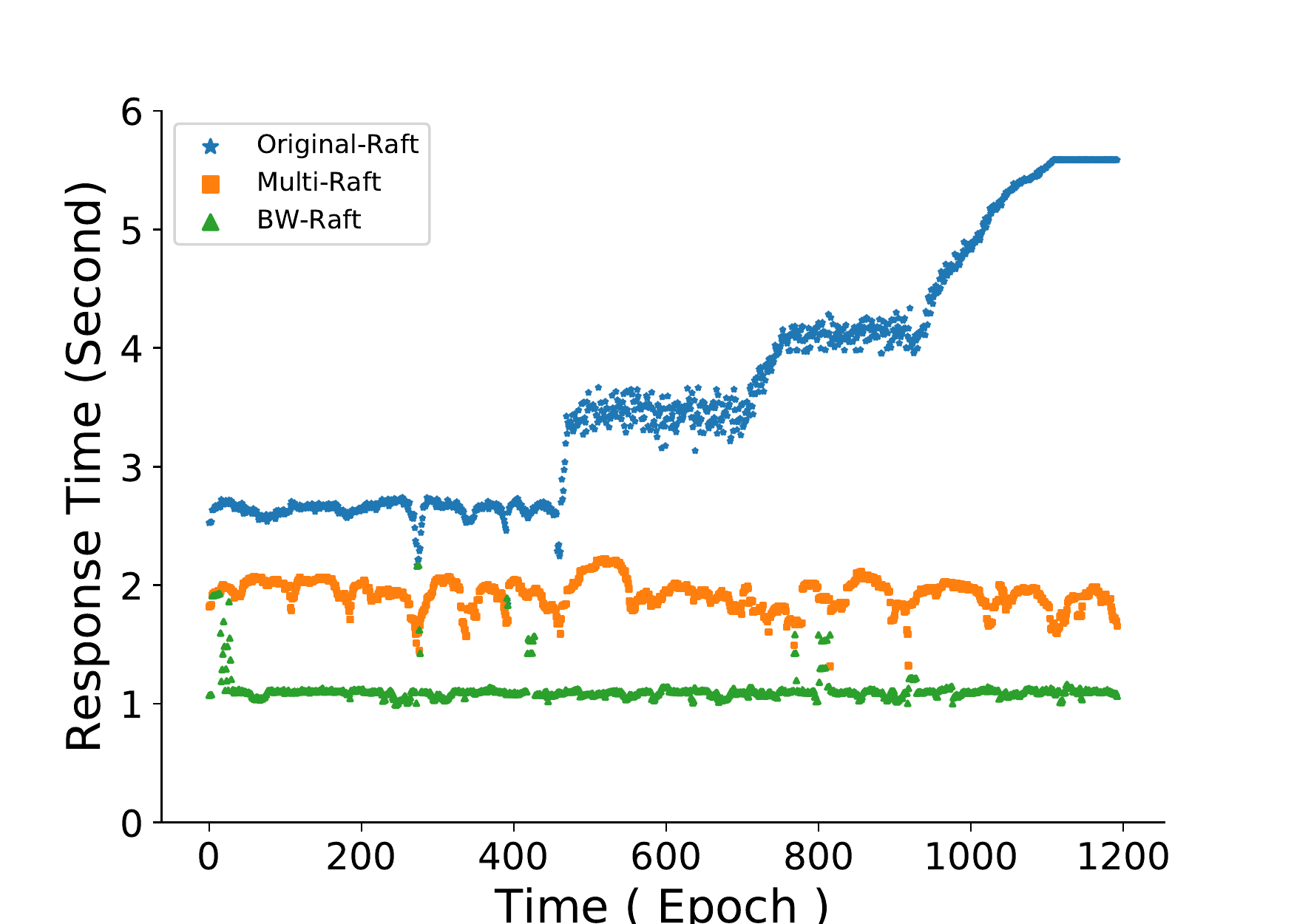} \\
\mbox{(a) Small} & \mbox{(b) Medium} & \mbox{(c) Large} \\
\end{array}$
\caption{50 days Performance snapshots running \textbf{Read} (top) and \textbf{Write} (bottom) workloads. Y axis is in log scale.}\label{f:wr}
\end{figure*}

%BW-Raft  realizes several interfaces which enable a scalable geo-diverse replicated state machine.  First, BW-Raft  provides an interface that implements strong consensus  between  sites. Among all sites, BW-Raft  leases global and local spot instances as secretaries for leaders. Within a site, BW-Raft hires local spot instances as observers for followers.  The total number of spot instances is dynamically tuned based on current performance statistics and resource provision, as discussed in \SEC{sub:modeling}. BW-Raft  also exposes the management interface  to clients, supporting upper-level applications. Hence, developers could switch to BW-Raft with a few extra lines of codes.

\subsection{Secretary Implementation }
%When the number of servers in a BW-Raft cluster is too large, leader needs to hire secretary to offload log replication task. 
Because BW-Raft is a consistent protocol targeting global service. Secretary could be deployed in worldwide data centers. The location of secretary could be selected specially to get lower latency, lower network cost, and higher throughput. Figure~\ref{f:sec} shows an example of BW-Raft how to hire a local spot instance as secretary to offload leader's log replication. When a leader hires a secretary, the leader specifics some followers to secretary and lets the secretary replicate log to the specified followers. And followers replies to secretaries whom inform the leader when the log has been appended to the majority of followers. BW-Raft can also hire global secretary to offload the communication task of confirming it's leader status. Due to spot instance can be interrupted at  any time, all the heartbeat messages are sent by leader to maintain leadership. If there is no secretary in the running time, leader directly replicates log entries to followers. In this case, BW-Raft reverts and equals a classic Raft. Because different secretaries number might result in a large gap between some leader and followers nodes. We also provide an interface to change the number of secretaries in a cluster.

\subsection{Observer Implementation }
% When the number of servers in BW-raft cluster is too small and needs to handle a large number of read operations.  
In order to deal with a large number of hot data access, we implement an interface to make BW-Raft can dynamically rent the spot instance as observer to offload client's read operation. Figure~\ref{f:obs} shows an example of BW-Raft how to hire a local spot instance as observer to offload client's read operation.  In BW-Raft, if observer receives a read request, it will request the newest \emph{readindex} from the leader, and return to the client after the state machine executes to readindex. However, with no additional measures, readindex operations  would run the risk of returning stale data, since the leader responding to the request might have been superseded by a newer leader of which it is unaware. In original Raft, before the leader response client's read request, the leader must send heartbeat to check if it still is the leader. But in BW-Raft, we can hire a global secretary to do this. Therefore, as long as the leader communicates with global secretary , it  knows whether it has been replaced or not.

\section{Evaluation}\label{SEC:EVA}

\begin{figure*}[t]
\centering
$\begin{array} {cc}
\includegraphics[width=3.1in]{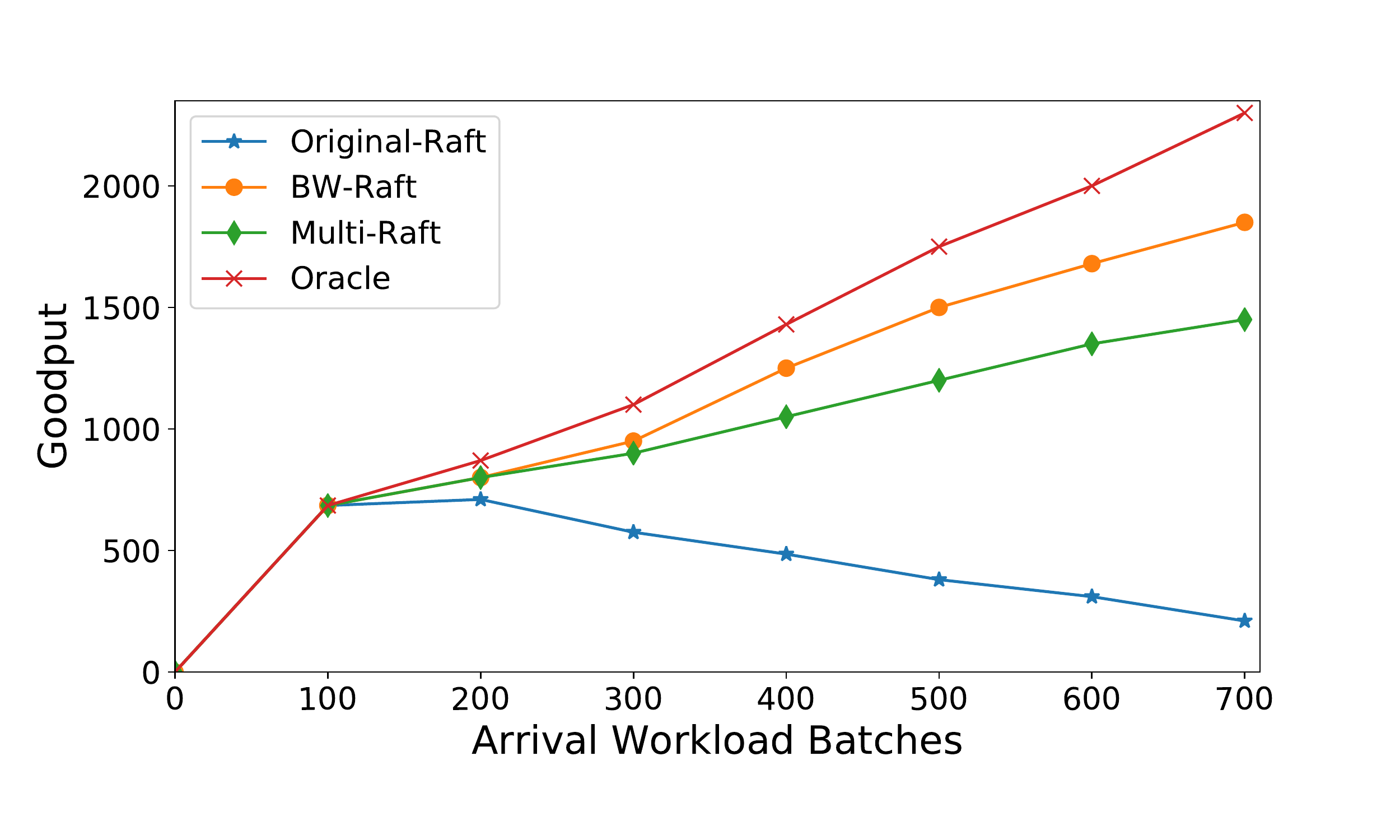}
&\includegraphics[width=3.1in]{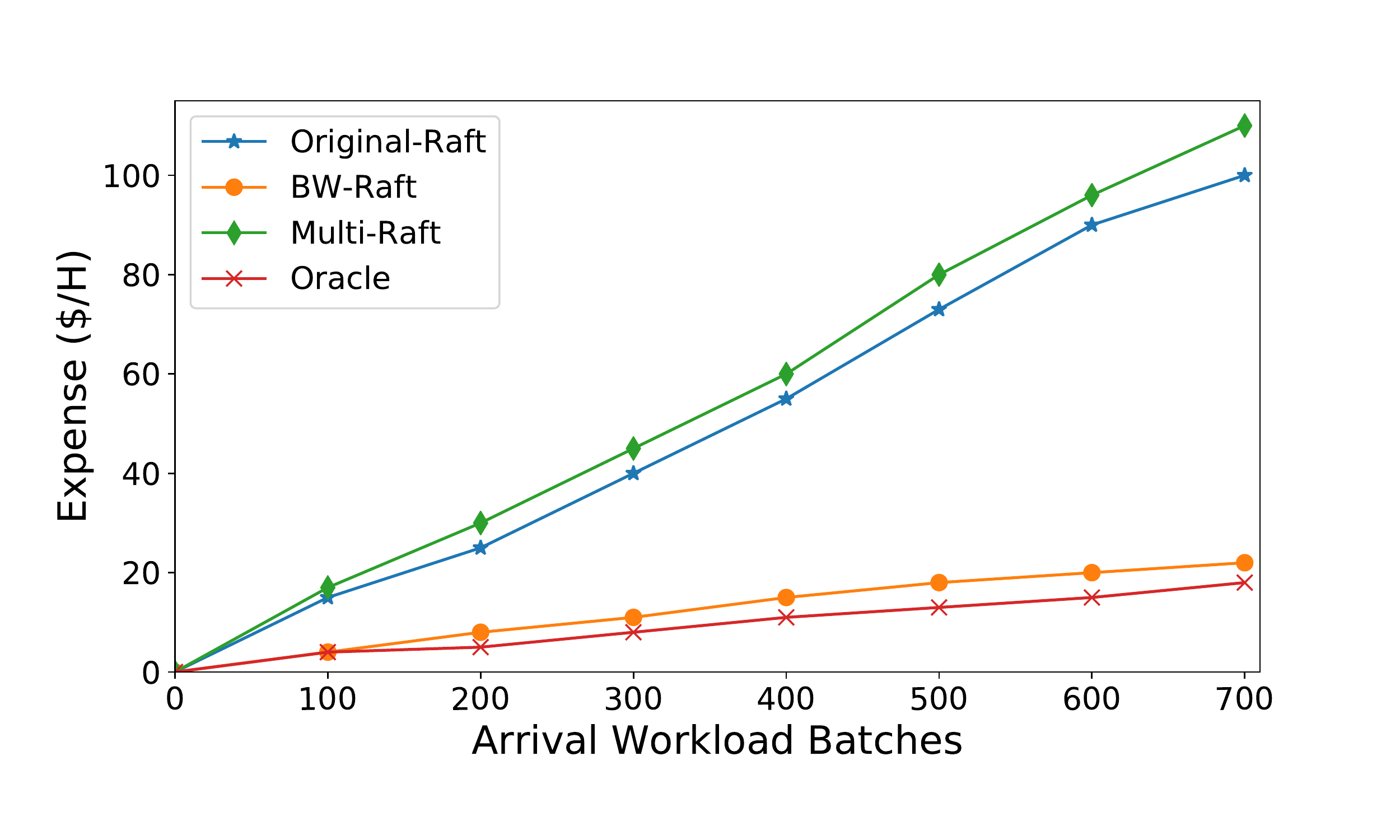} \\
\mbox{(a) Goodput} &\mbox{(b) Expense}
\end{array}$
\caption{Performance (a) and cost (b) of BW-Raft and other baselines at scale-out.}
\label{f:scale}
\end{figure*}

\begin{figure}
\centering
$\begin{array} {cc}
\hspace{-6pt}\includegraphics[totalheight=60mm,width=45mm]{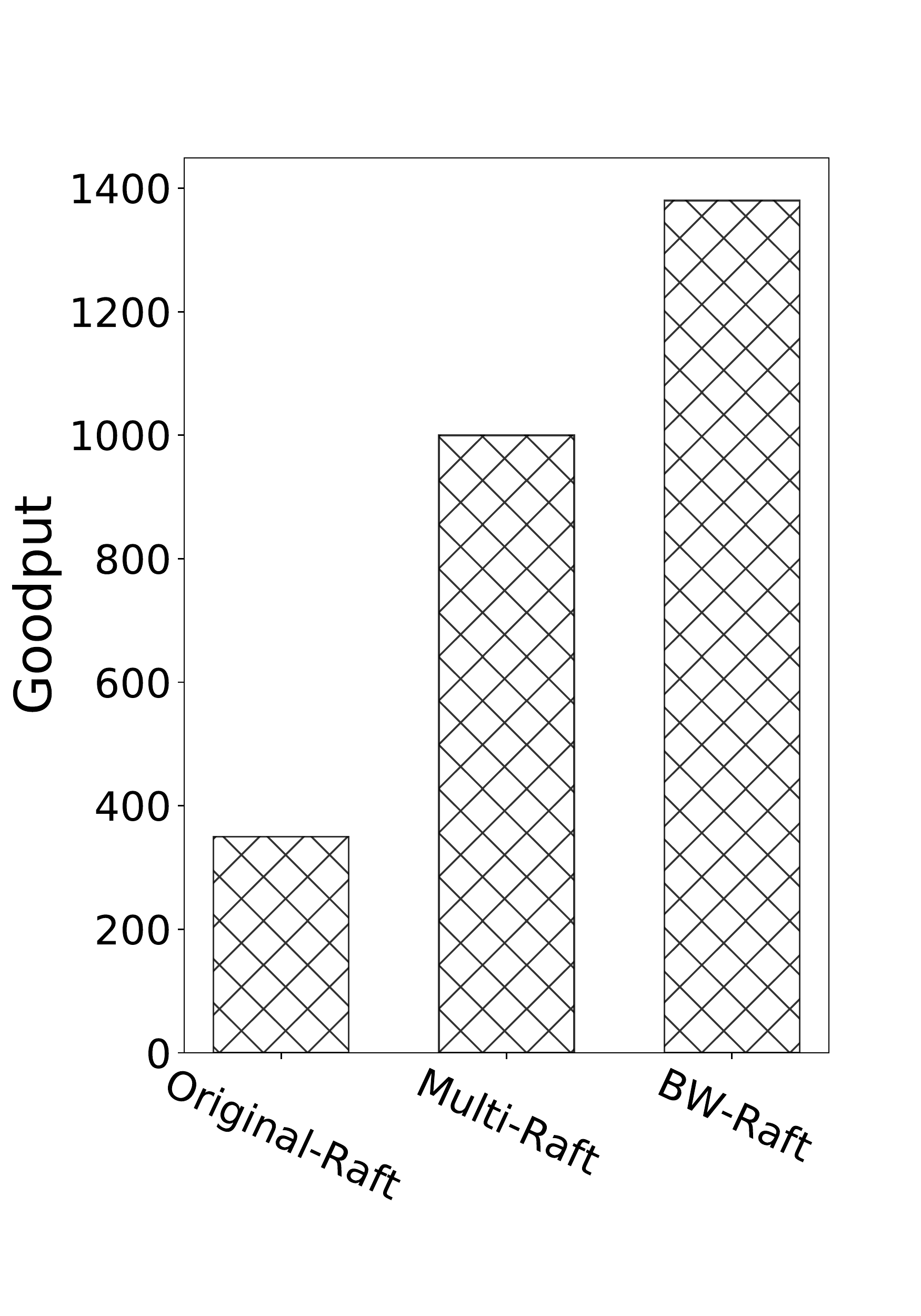} &
\hspace{-20pt}\includegraphics[totalheight=60mm,width=45mm]{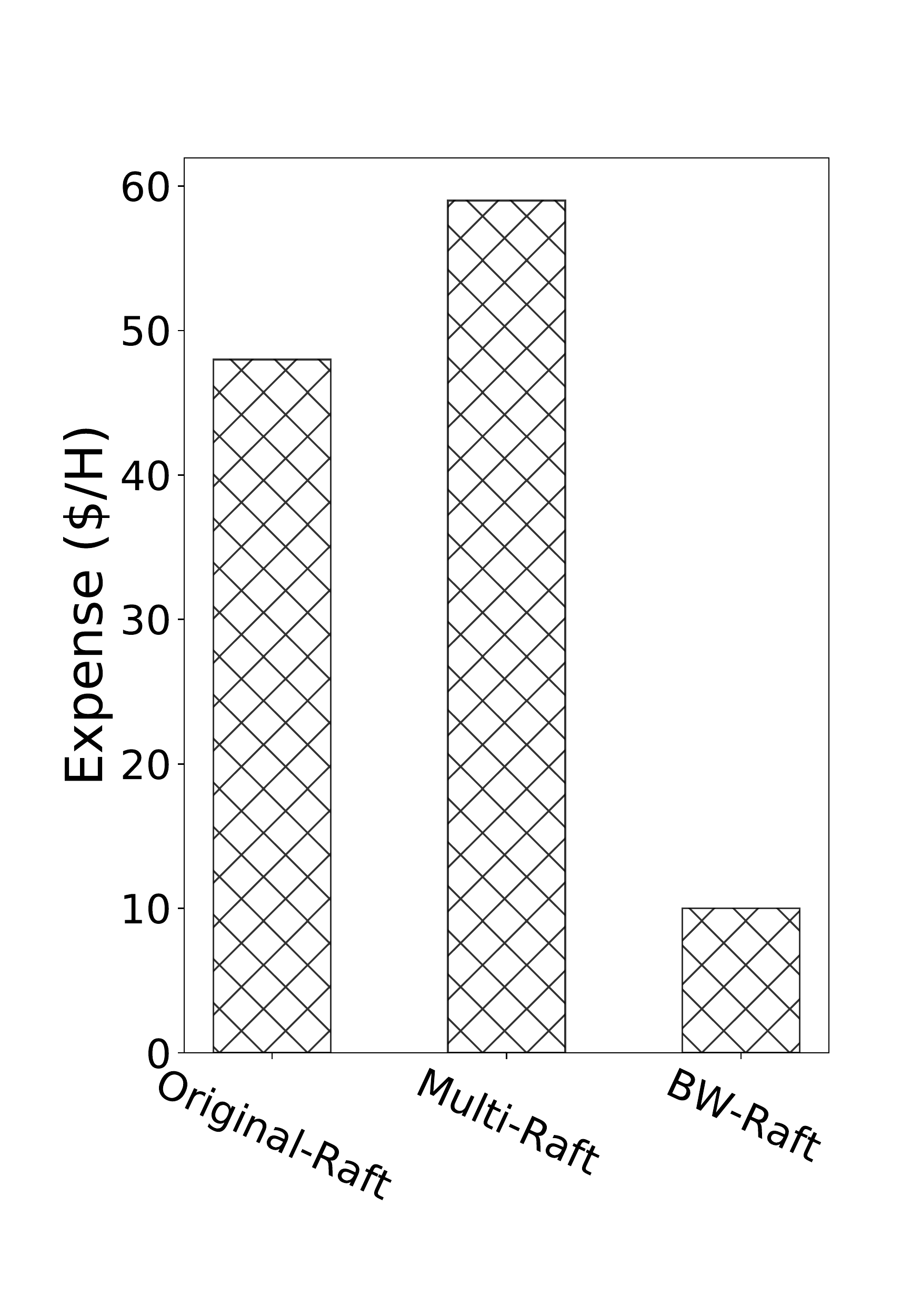} \\
\vspace{0pt}\mbox{(a) Goodput} & \hspace{0pt}\mbox{(b) Expense}
\end{array}$
\caption{Performance (a) and cost (b) comparison between BW-Raft and other baselines.}
\label{f:stat}
\end{figure}

\begin{comment}

\begin{table}[!htbp]
\centering
\caption{Instance Profiles of Testbed}\label{tab:ycsb}

\begin{tabular}{lc}
\toprule
System & Configuration\\
\midrule
Operating System & Ubuntu16.04 kernel 4.15.0  \\
CPU  & Intel Xeon 2603 1.70GHz  \\
Memory & 32GB DDR4 \\
\bottomrule
\end{tabular}
\end{table}

\end{comment}

BW-Raft is designed to achieve a cheap and scale-out consistency protocol by using on-demand instances and spot instances, which can handle different workflows and run on some unstable machines. 
In this section, we evaluate the performance of
BW-Raft by prototyping it onto our physical
testbed and Amazon AWS EC2~\cite{awsec2,morris2018model}.
%The instance profiles of our testbed are listed in Table~\ref{t:machstats}.

\noindent{\bf System Setup}: We deploy BW-Raft onto Amazon AWS EC2's instances (t2.small) to evaluate the runtime performance of BW-Raft by using Google workload. For the burstable spot instance, we report 0.415\$/H on average. Spot instances are up to 90\% cheaper than corresponding regular on-demand instances. We also build a physical testbed to illustrate the performance of BW-Raft by using YCSB  \cite{cooper2010benchmarking} workload. The testbed contains 12 servers with Intel Xeon 2603 1.70GHz and 32GB DDR4 memory, 2TB hhd and connected by switch router Ruijie RG-S2952G. 

%\subsection{Setup}\label{sub:set}
\noindent{\bf Software Setup}: The operating system is built as Ubuntu16.04  (kernel version  4.15.0 ) and the version of YCSB is 0.17.0. Our experiments are built based on a client-server model, as shown in Figure~\ref{f:archi}. The client sends batched workload as reads/writes operation, with different CPU/memory demand, arrived in a Poisson distribution. 

In this paper, we mainly evaluate BW-Raft with the following baselines:
\begin{itemize}
\item {\em Original} implements a state-of-the-art Raft design from Ongaro et al.~\cite{ongaro2014search};
\item {\em Multi-Raft} is a state-of-the-art multi-raft implementation using sharding~\cite{ho2016fast};
\item {\em Oracle} is a base line based on offline analysis.
\end{itemize}

\noindent{\bf Workloads and Traces}: We verify the performance of BW-Raft using real world workloads and traces. We use the popular Google cluster trace~\cite{gcluster} which contains one-month job statistics in Google cluster. Workloads are random reads/writes, controlled ratio R/W batches, and read/write-only workloads. All workloads are tested with small, medium, and large block size, namely 256KB, 1024KB, and 2048KB, respectively. 
\begin{itemize}
\item {\bf Read} is a batch of workload with read-only queries.
\item {\bf Write} is a batch of workload with all writes.
\item {\bf $\alpha$-Static} is an $\alpha$ controlled workload batch with $\alpha$ as the read/write ratio.
%\item {\bf Random}: is a workload with random reads and writes generated.
\end{itemize}

\noindent\textbf{Performance Snapshots}: First, we report the performance snapshot of running \emph{BW-Raft}, \emph{Multi-Raft}, and
\emph{Original}. The whole experiment lasts for
1200 epochs (i.e., 50 days) in Amazon EC2. Figure~\ref{f:wr} plots the average latency when
executing \textbf{Read} (top) and \textbf{Write} (bottom). For reads, BW-Raft provides the shortest
average response time (i.e., 1.26s), which is $27\%$  of Multi-Raft (4.6s) and $15\%$ of Original (7.9s).
However, in some extreme cases, BW-Raft can perform badly, such as overshoots in epoch 410 and 578 in
Figure~\ref{f:wr}(a) top. Similar overshoot happens when BW-Raft executes larger reads. Such overshoots happen when
the majority of \emph{observers} failed and BW-Raft had to reschedule workloads to correspondent \emph{followers}. The overshoot
can be  mitigated when we spreading observers onto many sites instead of a few cheap ones, which is a tradeoff
between performance and cost.

For write operation, BW-Raft scales out in proportion to the ever-increasing updates, nicely. Multi-Raft also scales, however, with a price
of 3X larger response time due to maintaining the 2 pc commit between leaders. Original cannot handle constant updates when scaling-out.
When the size of appended logs increases to a certain limit in all nodes, Original fails at the leader blocking the overall write performance, slowing down 2.5X compared
to BW-Raft. Overall, BW-Raft shows significant performance improvement, as it scales in increments 3-12X compared to Multi-Raft and Original.

\begin{figure}
  \centering
  \includegraphics[width=3in]{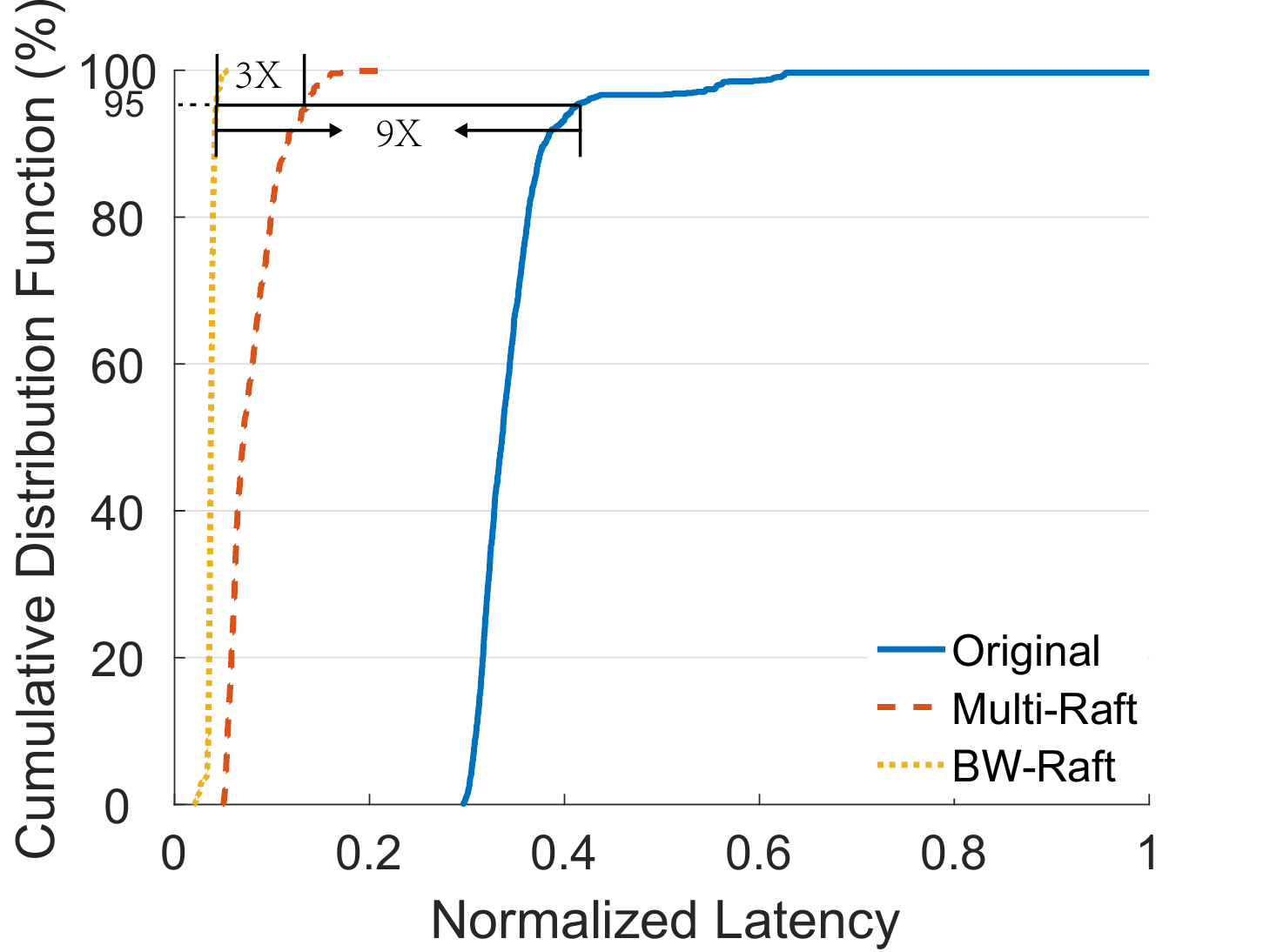}
  \caption{CDF of \gr~and baselines.}\label{f:CDF}
\end{figure}

\begin{figure*}[t]
\centering
$\begin{array} {cccc}
\includegraphics[height=0.22\textwidth]{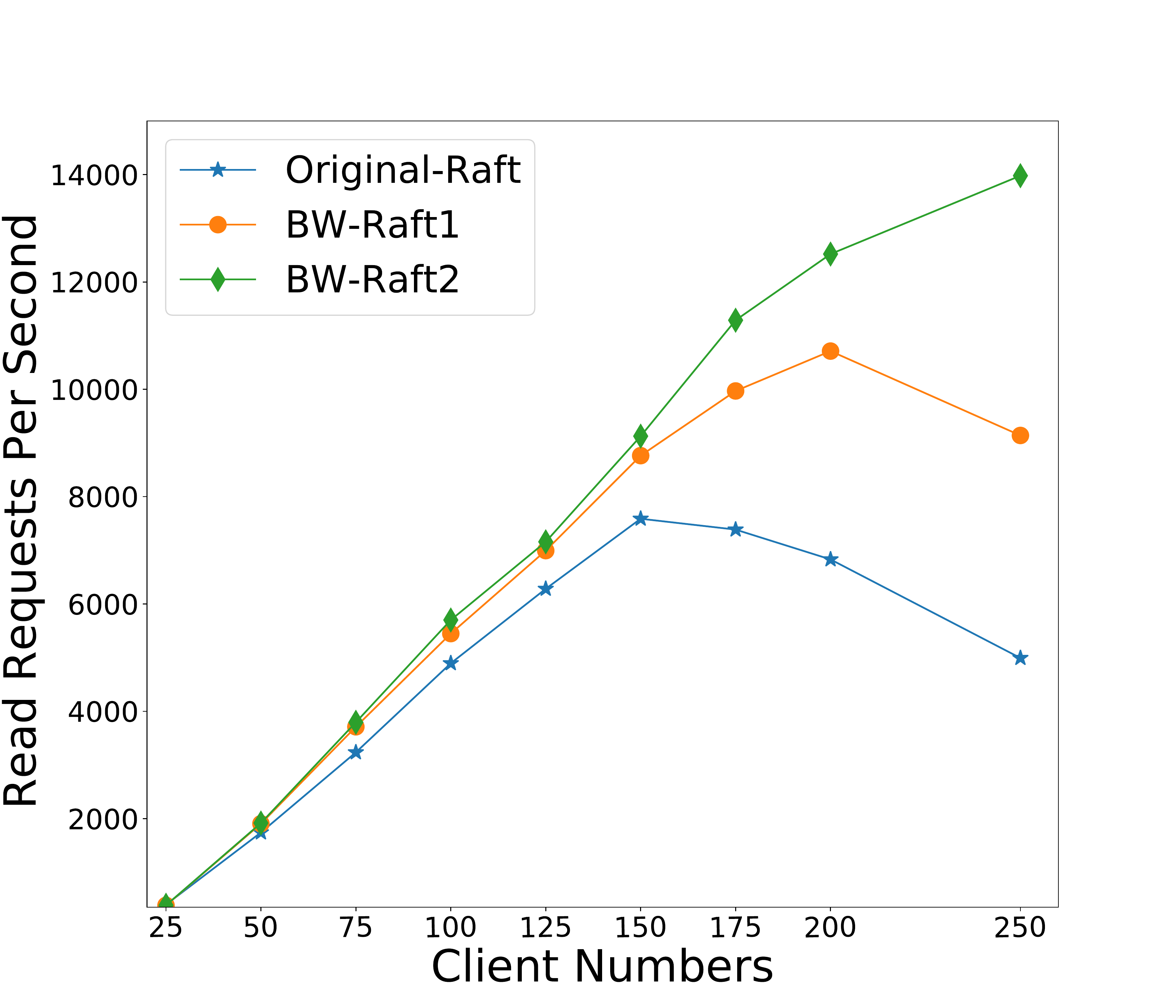} & \hspace{-12pt}\includegraphics[height=0.22\textwidth]{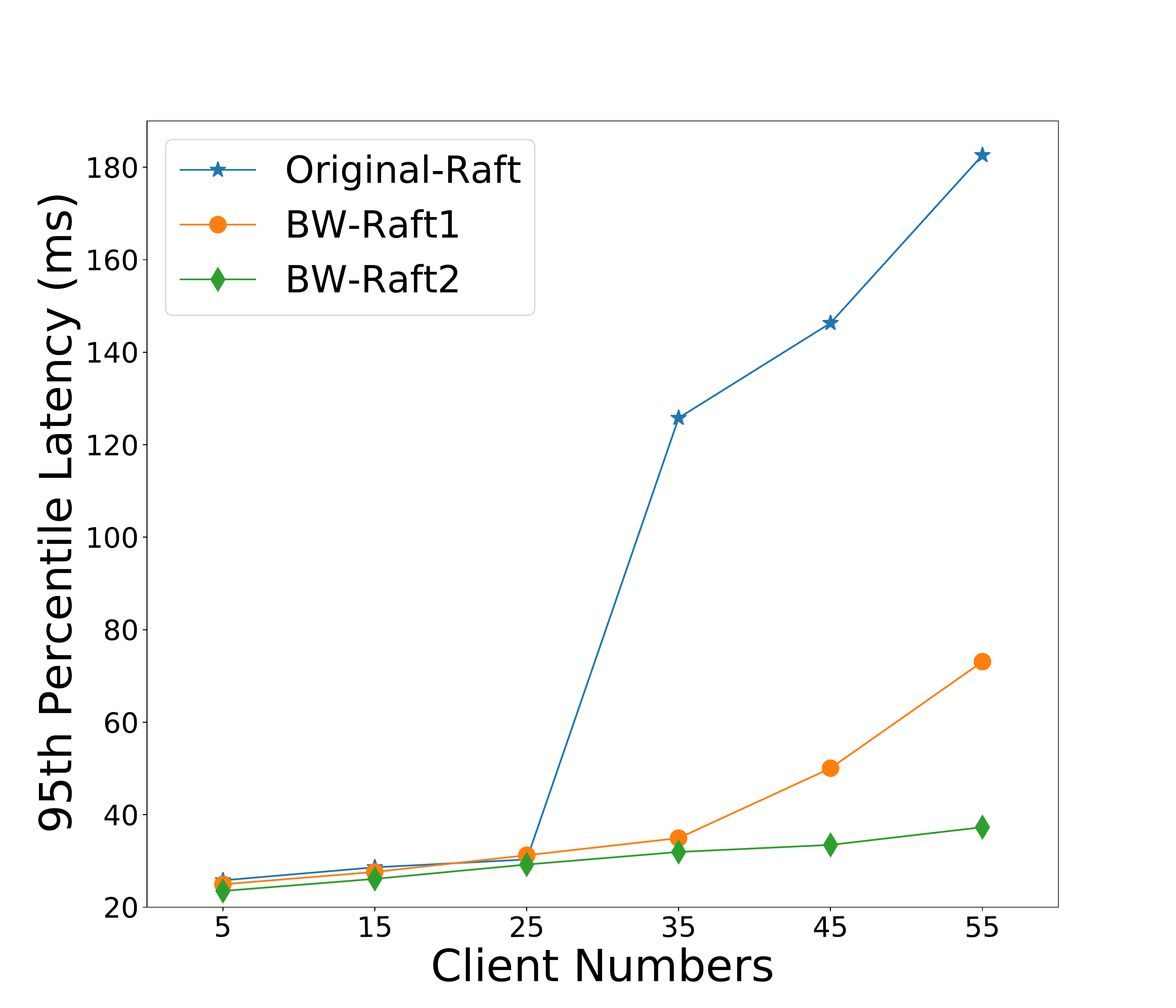} & \hspace{-12pt}\includegraphics[height=0.22\textwidth]{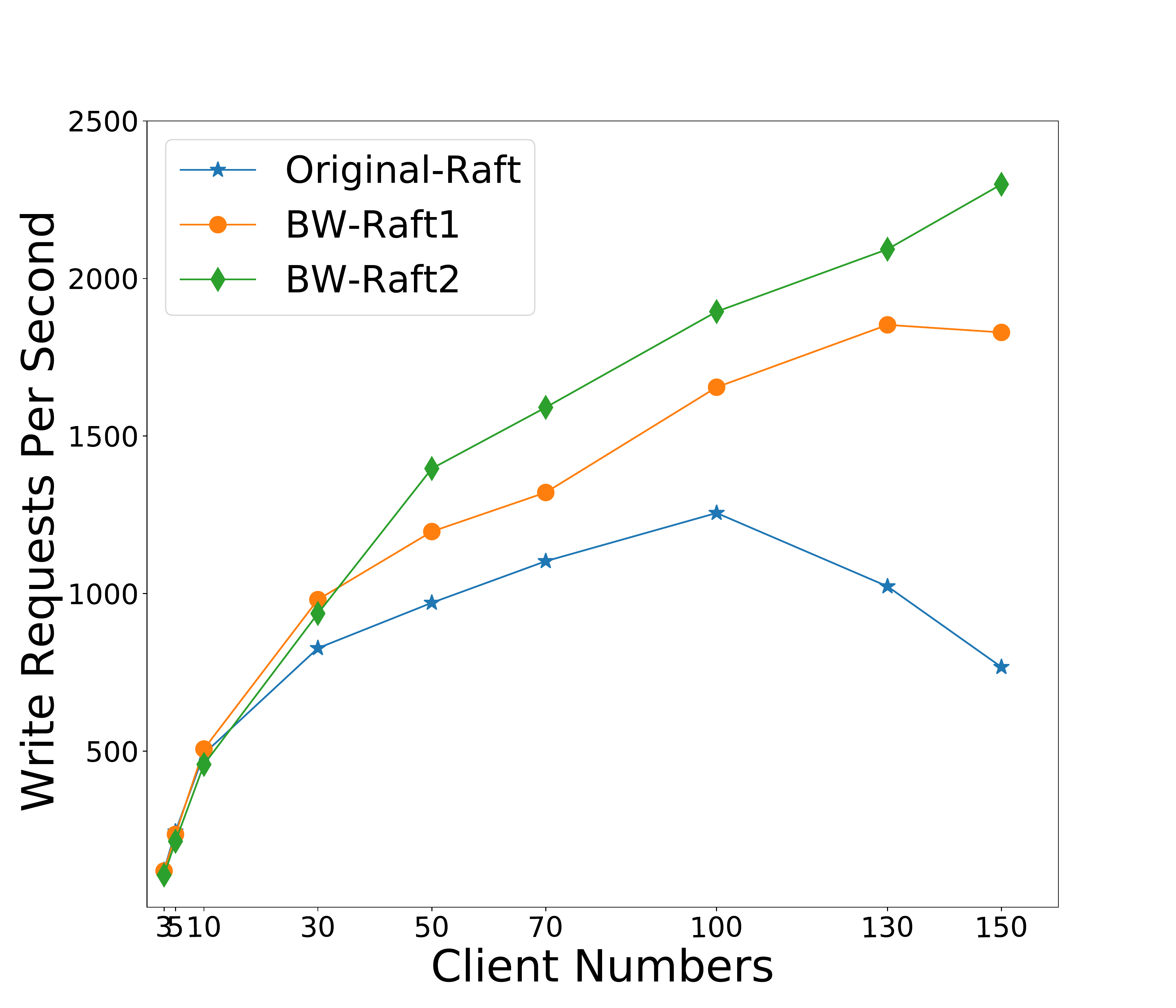} & \hspace{-12pt}\includegraphics[height=0.22\textwidth]{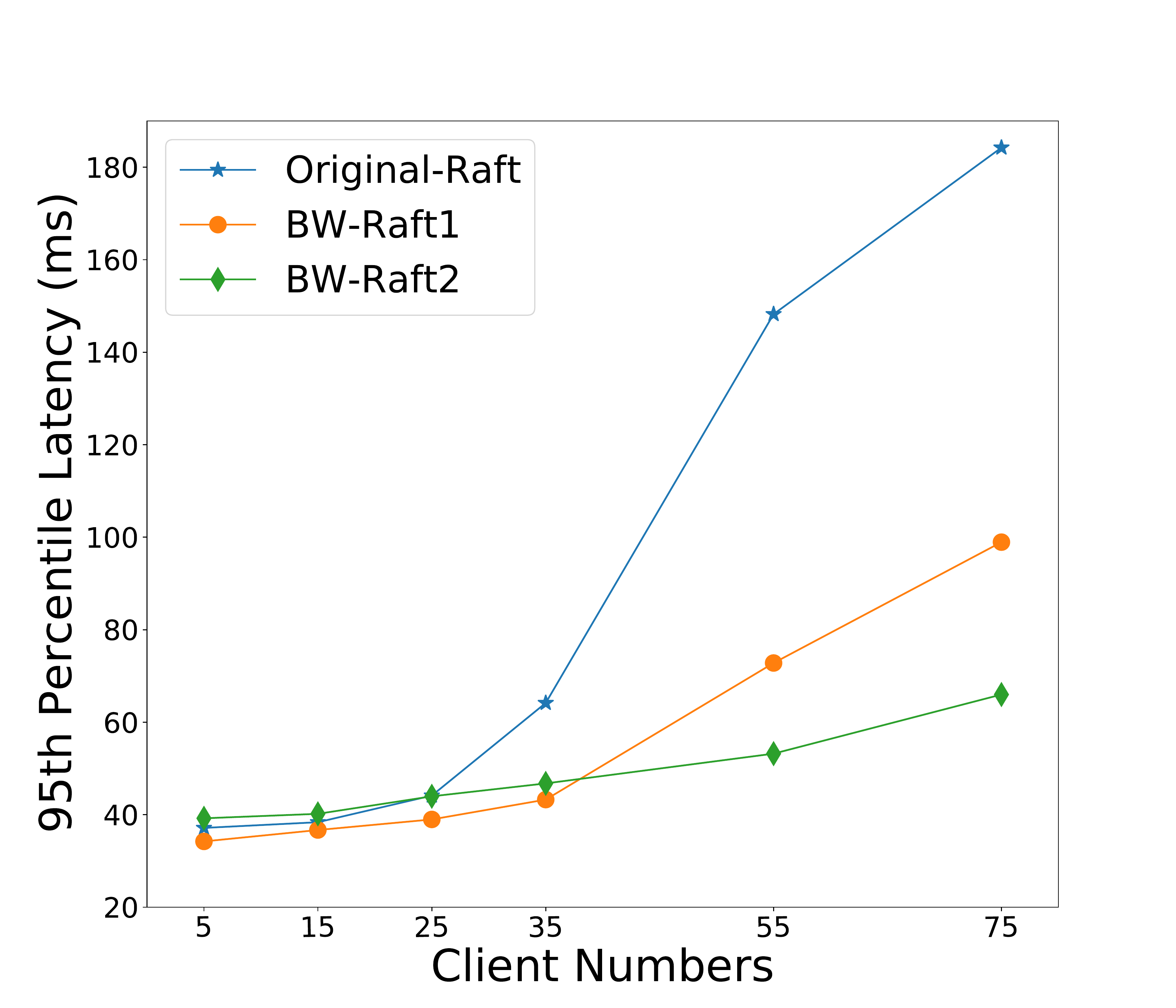}  \\
\mbox{(a) Read Goodput} & \mbox{(b) Read Latency} & \mbox{(c) Write Goodput} & \mbox{(d) Write Latency} \\
\end{array}$
\caption{Goodput and latency comparison between BW-Raft with different numbers of secretaries and observers. In figure(a)(b) BW-Raft1 and BW-Raft2 means BW-Raft with one observer and two observers.
In figure(c)(d) BW-Raft1 and BW-Raft2 means BW-Raft with one secretary and two secretaries. }\label{f:compare1}
\end{figure*}

\begin{figure*}[t]
\centering
$\begin{array} {ccc}
\hspace{-12pt}\includegraphics[height=0.28\textwidth]{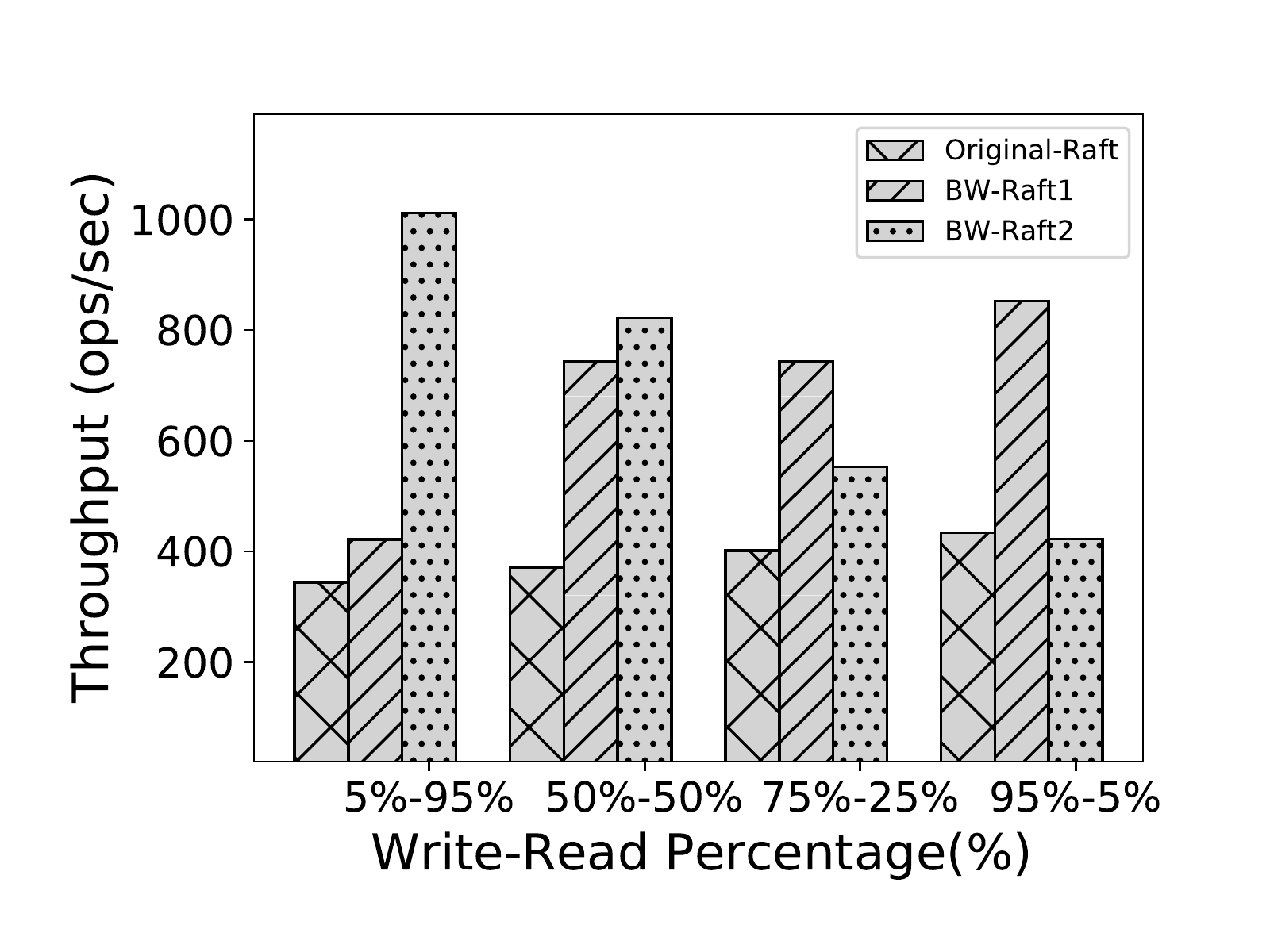} & \hspace{-27pt}\includegraphics[height=0.28\textwidth]{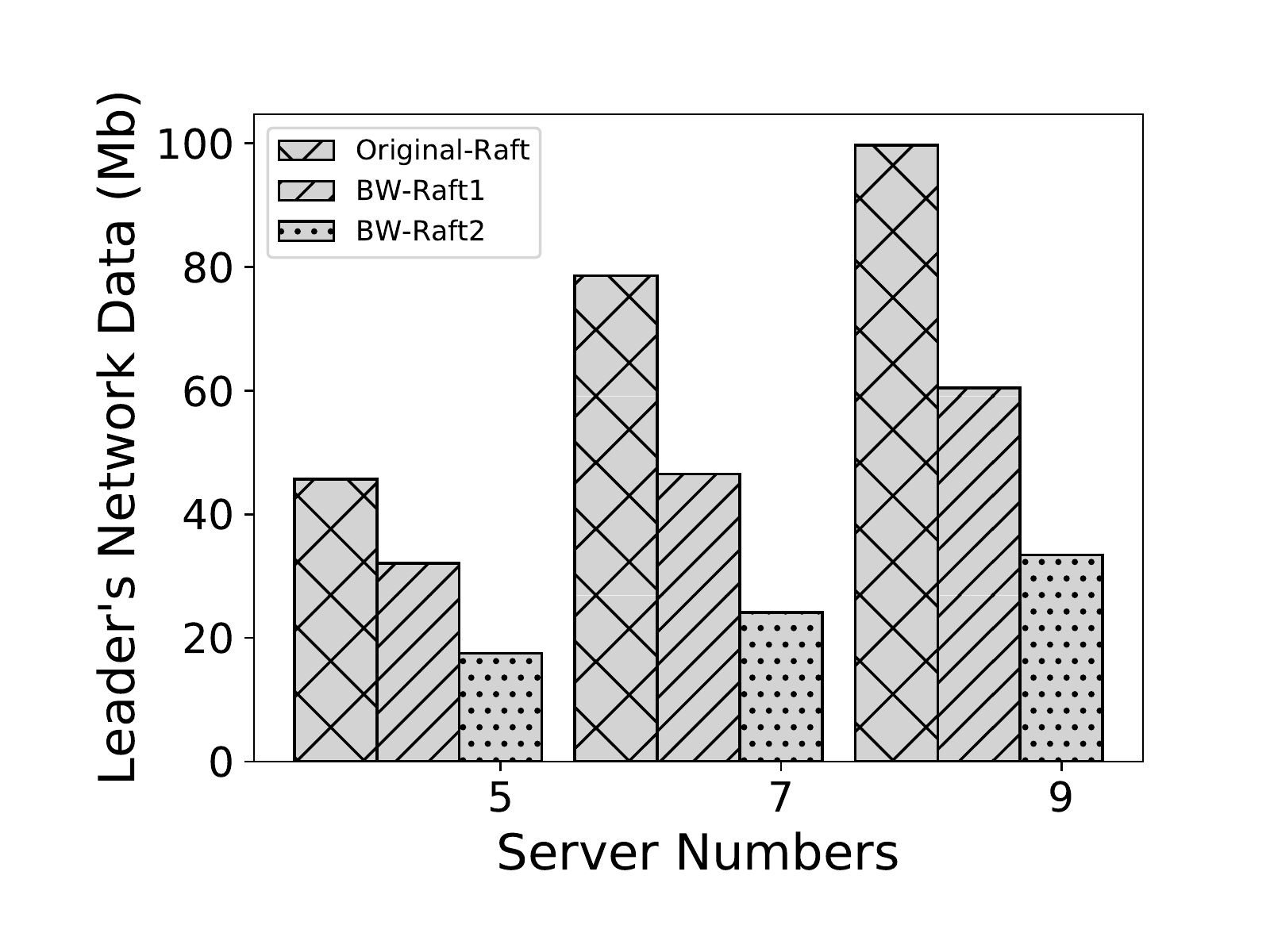} & \hspace{-27pt}\includegraphics[height=0.28\textwidth]{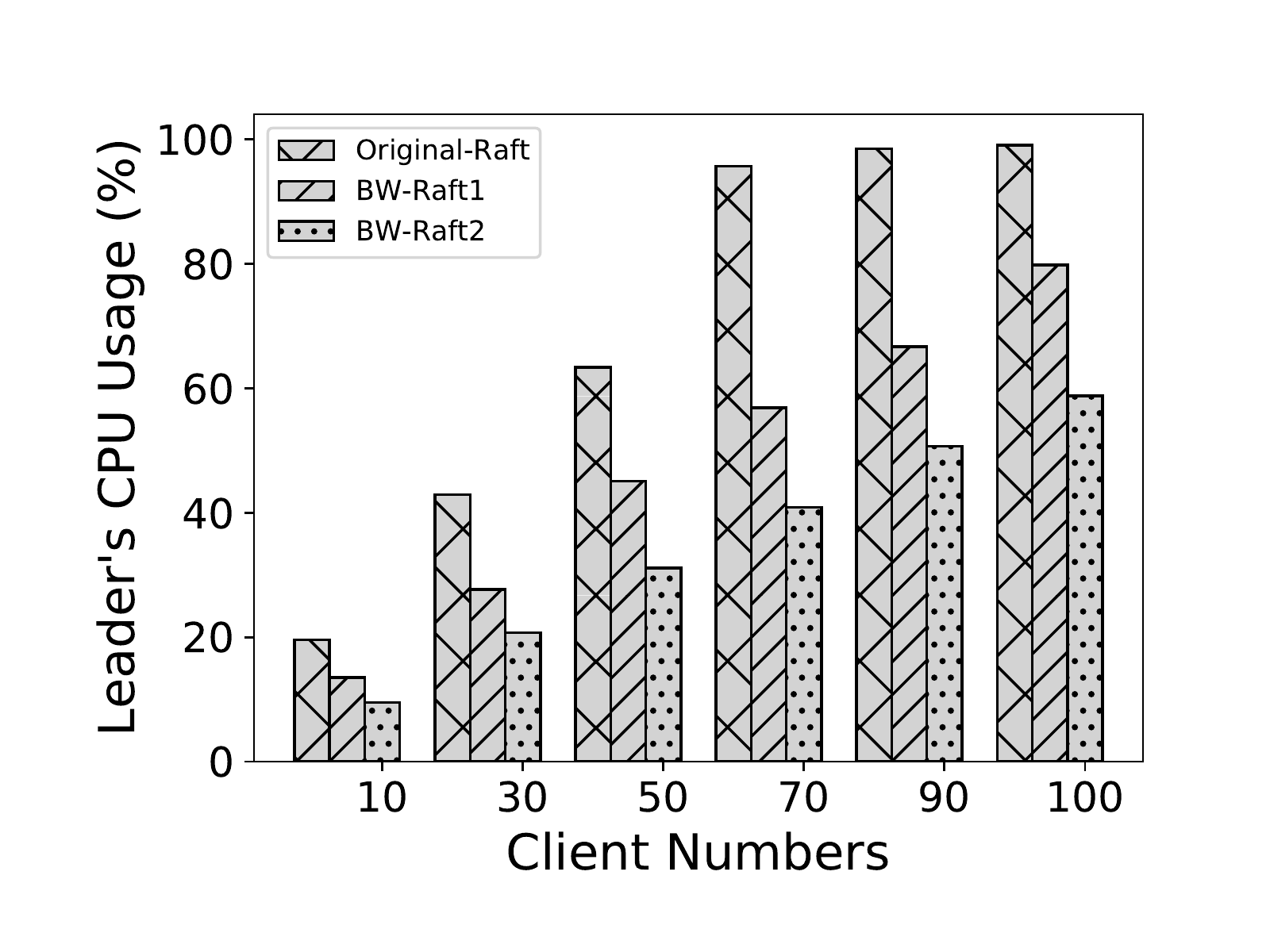}  \\
\mbox{(a) YCSB Benchmark} & \mbox{(b) Network Data Usage} & \mbox{(c) CPU Usage}  \\
\end{array}$
\caption{ Comparison of YCSB's benchmark result and leader resource usage after BW-Raft scale out. In figure(a) BW-Raft1 and BW-Raft2 means BW-Raft with one secretary and one observer.
In figure(b)(c) BW-Raft1 and BW-Raft2 means BW-Raft with one secretary and two secretaries.  }\label{f:compare2}
\end{figure*}

%\begin{figure}
%\centering
	%\includegraphics[width=3.1in]{figure/sec-m.pdf}
	
	%\vspace{-0.3in}\caption{Find the optimal value %of $m_{write}$ }\label{f:sec-m}
%\end{figure}

\noindent\textbf{Scalability}:
BW-Raft can easily scale  in proportional to performance and cost. Figure~\ref{f:scale} demonstrates such proportionality when the workload
batch size increases by 700X. In Figure~\ref{f:scale}(a), both BW-Raft and Multi-Raft grow in proportional to the workload size. BW-Raft exhibits a better scale-out performance than Multi-Raft, and is close to the theoretically best performance in Oracle. Original does not scale. When the workload increases, Original suffers from managing too many logs at the leader node, which prevents scale-out. Multi-Raft can scale out
at a much higher cost. As shown in Figure~\ref{f:scale}(b), Multi-Raft
leases more on-demand nodes in scaling out while BW-Raft harnesses the salient feature of cheap spot instances. Compared to Oracle, BW-Raft still has
rooms to improve, especially in the workload and resource provisioning.

%  \centering
%  \includegraphics[width=4.1in]{figs/.pdf}
%  \caption{Throughput  under different workloads.}\label{f:scala}
%\end{figure}

%============================
\noindent\textbf{Overall Statistics}: Figure~\ref{f:stat} shows the overall performance and expense comparison between BW-Raft and other baselines. In goodput (i.e., correct queried result over unit time), BW-Raft is 7X and 1.5X, larger than
Original and Multi-Raft, respectively. Considering both reads and writes, BW-Raft has smaller variation than Multi-Raft. BW-Raft has significantly smoothed write delay curve due to secretaries often reside in more sites than \emph{followers}, which reduces unexpected long  wide-area network delay. For expenses, BW-Raft exploits cheap spot instances for secretaries and observers in many sites.
BW-Raft spends  86\%  and 80\% less than Multi-Raft and Original, respectively. Multi-Raft  usually costs more than Original due to its multiple
leaders thus expensive resource footprints. Note that, it seems unfair to show this expense comparison while only BW-Raft can use spot instances. However, there is no existing design on both Raft and Multi-Raft that can exploit spot instances. If we deploy spot instance in Original and Multi-Raft, they can be drown in a nonstop leader re-election and provide barely zero performance.

\begin{figure}[t]
  \centering
  \hspace{-12pt}\includegraphics[width=3.5in]{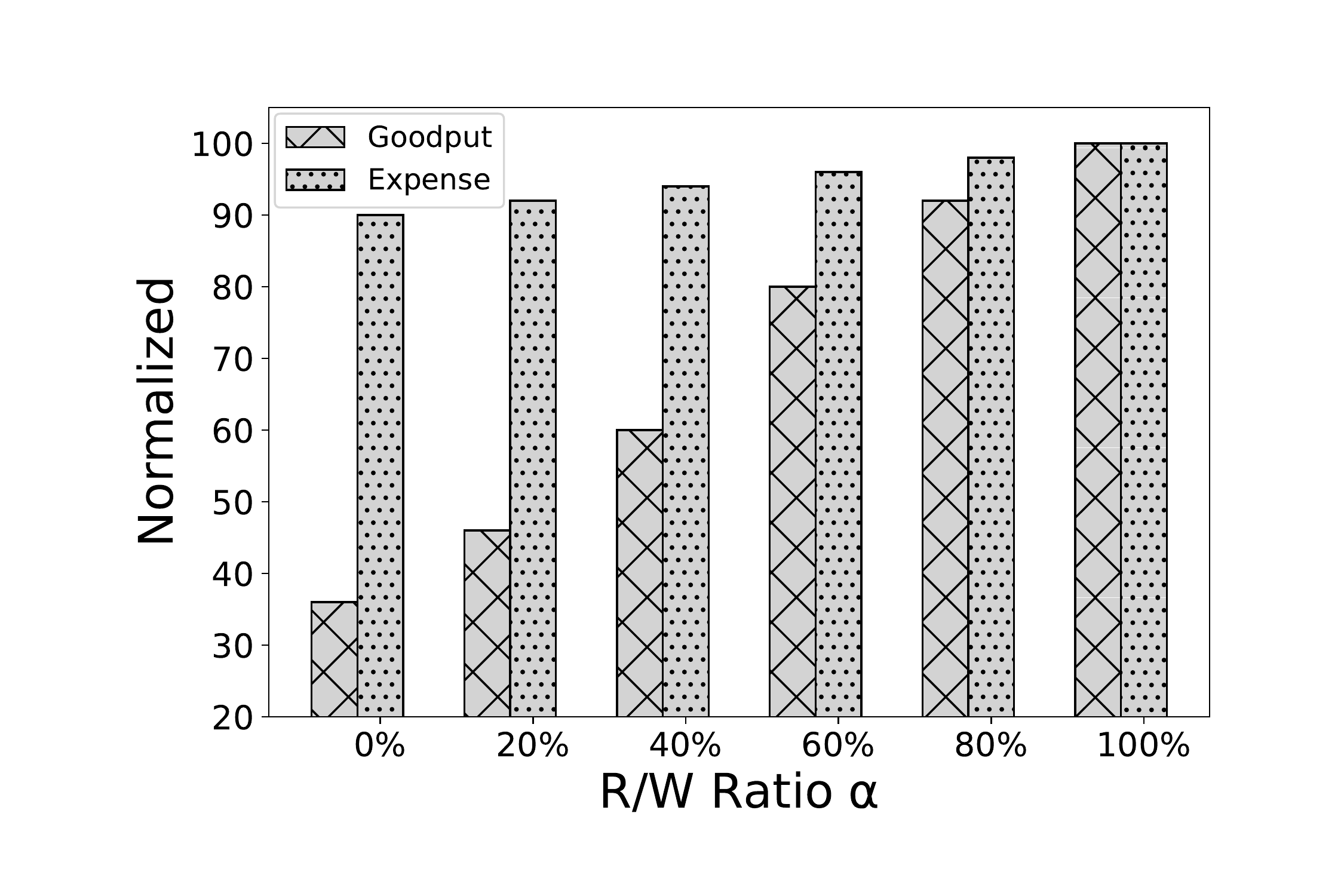}
  \caption{The impact of R/W ratio $\alpha$ in BW-Raft.}\label{f:alpha}
\end{figure}

%============================
\noindent{\bf Performance Distribution}. Figure~\ref{f:CDF} demonstrates the curriculum distribution functions of all jobs
running in BW-Raft and other baselines. A small portion of jobs do suffer in BW-Raft due to runtime spot instance failures and
errors in resource provisioning. In the worst case, BW-Raft could shrink back as a Raft handling a small number of jobs. However, BW-Raft has a much shorter tail than Multi-Raft in the latency distribution.
Comparing the 95th-percentile SLO, BW-Raft performs 3X better than Multi-Raft, and 9X better than Original.

\begin{figure}
  \centering
  \includegraphics[width=3.1in]{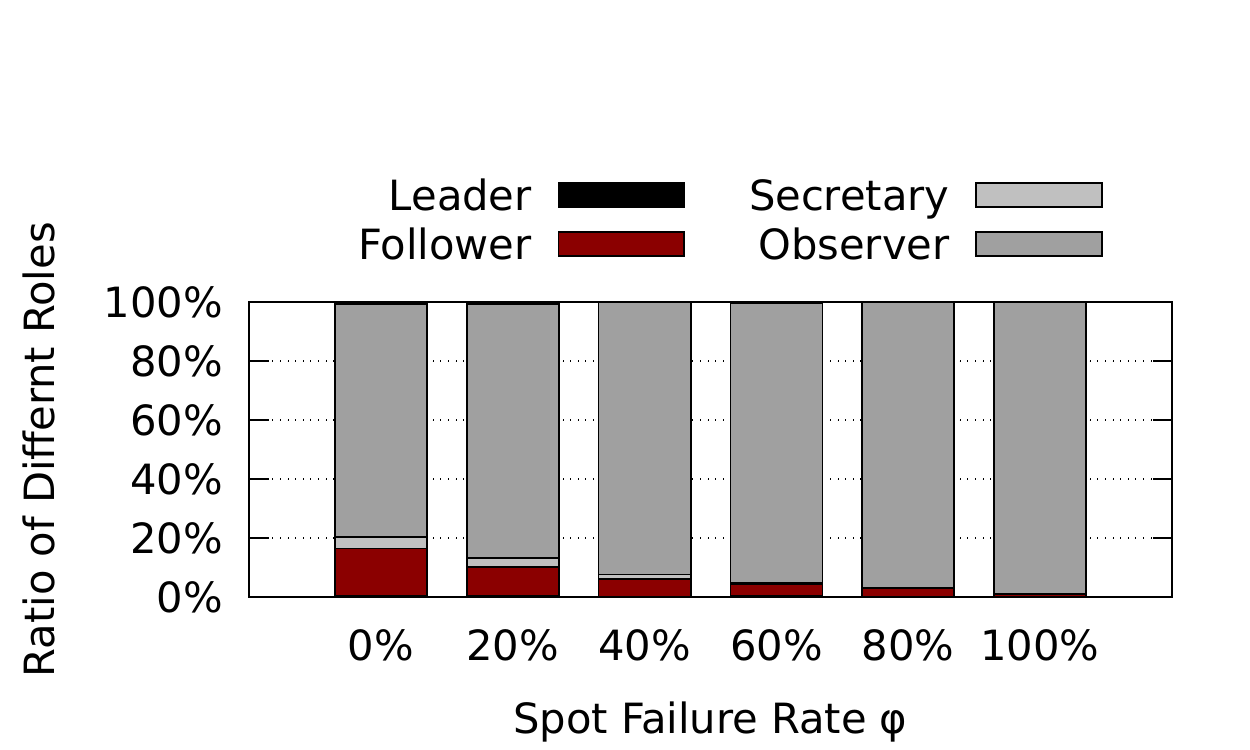}
  \caption{The impact of spot failure rate $\phi$ in \gr.}\label{f:phi}
\end{figure}

\noindent\textbf{BW-Raft's Performance with Different Configuration}:
Because BW-Raft increases the 
role of secretary and observer compared to original raft, which used to offload the read and write pressure. In Figure~\ref{f:compare2}(a), the different YCSB benchmark tests show that BW-Raft’s average throughput is 1.5-2 times than original Raft. However, as Figure 9 shown, the improvement of reading and writing performance is not significant when the number of clients is small. Besides, adding secretary can even lead to the performance degradation as shown in Figure~\ref{f:compare1}(d), due to the communication latency between the leader and the secretary. But with the increasing number of
clients, more and more network connections and data need to be processed, and the CPU resources of Raft leader will soon be exhausted as shown in the Figure~\ref{f:compare2}(c). So the Raft’s response delay will become higher and higher. Therefore, appropriately increasing the number of secretary can reduce the write latency and improve. Besides, because part of the log replication tasks are transferred from the leader to the local secretary, the network bandwidth of the leader is greatly reduced, which can greatly improve the performance in the limited network environment as Figure~\ref{f:compare2}(c) shown. In the case of read-only workload,  appropriately increasing the number of observers can greatly improve the throughput and reduce the response time as Figure~\ref{f:compare1}(a)(b) shown.

%============================
\noindent\textbf{Impact of Design Factors}: In our provision process, we have two major factors (i.e., workload R/W ratio $\alpha$
and spot instance failure rate $\phi$). The   factor $\alpha$ affects the provision process in BW-Raft, while the   factor $\phi$
could collapse the provision decision. Figure~\ref{f:alpha}
illustrates the impact of $\alpha$. The average goodput is increasing linearly
when BW-Raft serves more reads than writes. BW-Raft can handle reads well because BW-Raft
abusively employs many observers to serve reads.  These observers are   cheap, thus the overall expense in BW-Raft grows much slower than performance gain.
\begin{figure}
  \centering
  \hspace{-12pt}\includegraphics[width=3.1in]{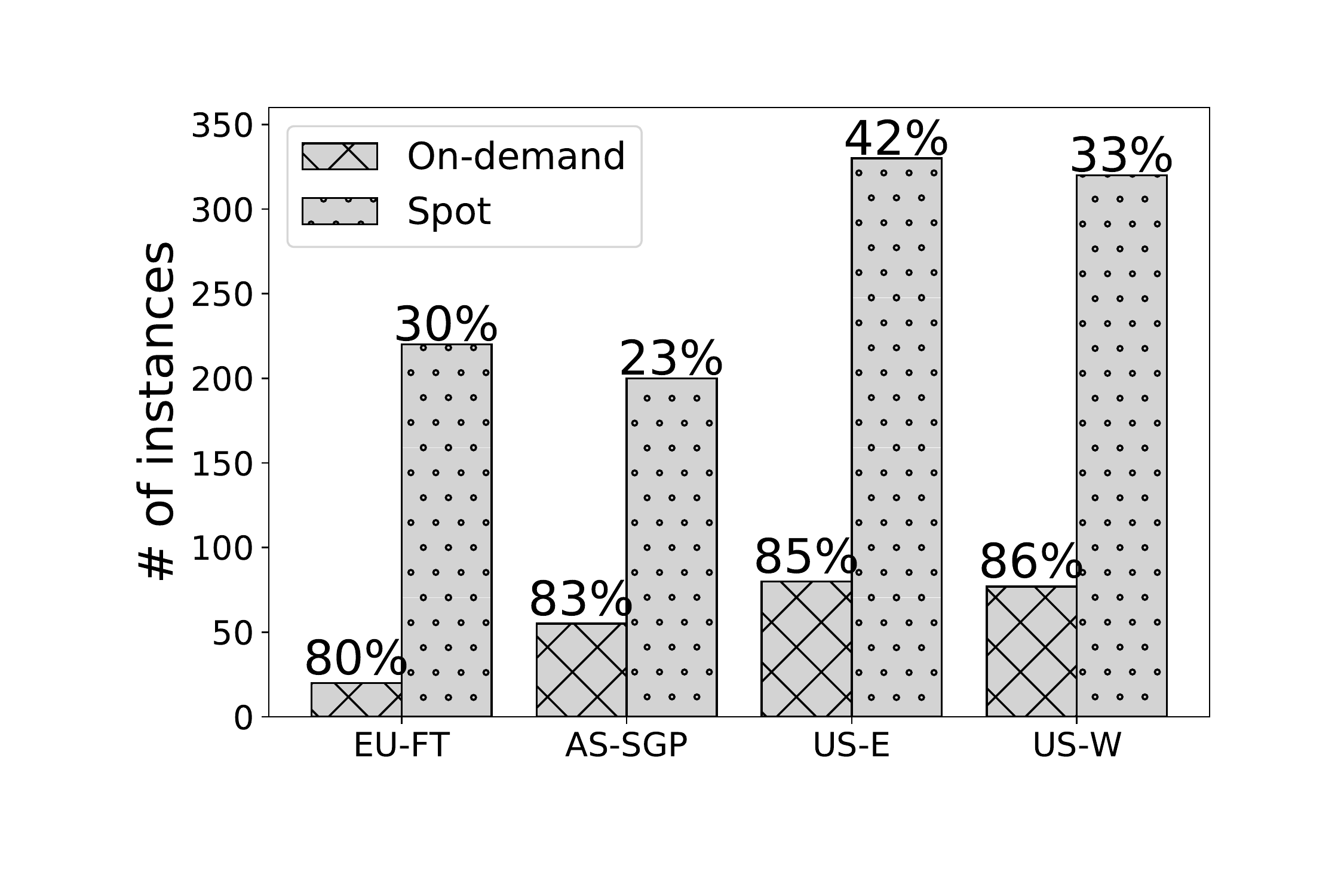}
  \vspace{-0.3in}\caption{Number of leased instances and average utilizations (on top of the bar) of BW-Raft in EU-Frankfort, Asia-Singapore, US-East, and US-West.}\label{f:mach}
\end{figure}

%\textbf{Impact of Leased Secretaries}:
Figure~\ref{f:phi} shows how spot instances fail affects BW-Raft. In our model, we assume a static $\phi$ based on history analysis, however, $\phi$
is hard to predict at runtime.
When this failure rate increases in a single site, BW-Raft gradually reduces the number of secretaries
at this site, and increases the number of observers in other sites. As such, the number of
followers decreases since secretaries decrease. BW-Raft reduces secretaries on purpose, in order
to reduce the possibility of update failure, and thus the cost from the consensus management.
While BW-Raft suffers  performance loss
from writes, it hedges  performance gain from reads as it hires more observers.

%============================
\noindent\textbf{Limitations of BW-Raft in the Wild}: We report the server-side statistics in Figure~\ref{f:mach}. We employ instances
from EU-Frankfort, Asia-Singapore, US-East, and US-West. BW-Raft hires more than 1500 instances, on-demand and spot, in these four sites. As shown in
Figure~\ref{f:mach}, BW-Raft hires 15-20X more spot instances than on-demand ones. On the one hand, when BW-Raft leases on-demand instance,
it takes the maximum use of the resource, exhibiting more than 80\% utilization in all sites. On the other hand, BW-Raft abusively leases spot instances without
fully exploiting its resources, with an average of 35\% utilization. It is mainly because of the failure-prone behavior of spot instances that BW-Raft only uses
a short period during their living period. If the burstable period of spot instance can be accurately predicted, BW-Raft can provide much better performance at scale-out. In addition, BW-Raft can only run in a non Byzantine environment. If there are malicious nodes involved, BW-Raft will not be able to guarantee the consistency of the system.

\section{Related Work}\label{SEC:REL}
Consensus algorithms, such as Paxos and Raft, are designed to  maintain consensus across multiple shared data replicas.
These algorithms scales by sharding~\cite{rao2011using} or automatically adding/dropping followers~\cite{deng2012adaptive,harlap2018tributary,wang2018effective,gramoli2016rollup}. However, for high write throughput, applications turn to over-provisioned sharding, multiplying inefficiency.
Our work, BW-Raft scales incrementally and does so using
cheap, failure-prone spot instances.
%In a distributed system, data is usually replicated in multiple data centers so as to  provide more efficient and service to the clients and enhance fault tolerance. In designing such a system, the consistency of the data should be placed at the primary position. While ensuring security and stability, we also minimize the cost of data storage and transmission, likewise maximize system performance.

\noindent\textbf{Raft Consensus Algorithm}: In the past few decades, researchers have proposed a large number of consensus algorithms \cite{ ongaro2014search, lamport2001paxos}. %Multiple nodes use consensus algorithm to ensure the consistency of data in non Byzantine \cite{Lamport1982The} situations and maintain availability even some machines fail.
Raft algorithm~\cite{ongaro2014search} is one of
the most widely used~\cite{tidb, cockroachlabs}.
Many companies have found that Raft is easy-to-implement
and provides good performance.
P{\^a}ris
et al. reduced energy footprint of
Raft~\cite{paris2015reducing}. Gramoli
et al. put
forward a fast consensus-based dynamic
reconfigurations method which can speedup a
primary-based rolling
upgrade\cite{gramoli2016rollup}. Copeland
et al. propose a Byzantine Fault Tolerant variant
of the Raft consensus algorithm
\cite{copeland2016tangaroa}.
%A modified
%dynamic-linear voting protocol was put forward to
%solve its high energy footprint
%\cite{Paris2016}.
These approaches have not explored efficient scaling out on Raft, while our work focusing
on improving the scale-out performance with Raft on
spot instances.

%Raft is a stronger leadership protocol. Cluster ensure consistency of data by selecting a unique leader.  In general, there are three roles in a cluster, leader candidate and follower. Leader maintains its leadership by constantly sending heartbeats to the followers. All data which is duplicated from leader to followers is also transmitted through these heartbeats. In addition, Raft also uses log matching and other security restrictions to force followers replicated identical to leader's. Every followers is watching a randomized election time out. If a follower do not receive broadcast from leader within election time out, this follower will increase its term number become a candidate and launch a new leader election. At the stage of leader election, every vote for a candidate whose log is not older than its. If a candidate gets the votes of most nodes, it will become a new leader. As mentioned above, leader election, log replication and some safety restrictions make up the basic Raft algorithm.

\noindent\textbf{Geo-Replication}:  Droopy and
Dripple \cite{Liu2017} , two sister approaches,
reduce latency by dynamically reconfigure leader
set. Tuba \cite{Ardekani2014} improves utility by
automatic reconfiguration. SPANStore \cite{Wu2013}
offers low cost storage services making use of the
price difference between
suppliers. Cadre~\cite{xu2015cadre},
Lynx~\cite{Zhang2013}, and Flutter~\cite{hu2018time}
achieve low latency by avoiding long distance
transmission.

\noindent\textbf{Spot Instance Market}: In the
cloud market, suppliers provide on demand instance and spot
instance.  Spot instance is usually much
cheaper than on-demand instance. However, due to
spot instance is unstable and may stop at any
moment, it is not reliable for data tasks, especially
in maintaining data
consistency. There are
many research reduce cost by using spot
instances. EAIC \cite{Jangjaimon2015Effective}
reduce cost by adaptive checkpointing. And PADB
\cite{Song2012} algorithm was put forward get
maximized mean profit.

\section{Conclusion}\label{SEC:CON}

In a cloud computing environment, it is important to support intensive data service at scale out. While preserving strong consistency, it is expensive and complex support this data-intensive services at geo-diverse sites.
In this paper, we proposed BW-Raft, an extended Raft framework that  offloads log management to secretaries and offloads read request to observers as a new scheme to support strong consistency between services. We designed a global resource management to make this management cheap. In practice, we prototype a key-value store service in BW-Raft framework. Our source code can be found at ~\cite{sourcecode}. We analyzed the performance of BW-Raft with many other  protocols, and concluded that that (1) BW-Raft significantly boosts throughput by up to 9X, compared to Raft, and (2) BW-Raft is 84\% cheaper than Multi-Raft. In general, BW-Raft is a practical framework to support scale out data-intensive computing across geo-diverse sites.

\section{Acknowledgments}\label{SEC:ACK}
This work was supported in part by Ministry of Science and Technology of China (Grant No. 2018YFB1404303), also the National Natural Science Foundation of China (Grant No. 61702250), and the Science and Technology Department of Jiangxi Province (Grant No. 012031379055).

%\bibliographystyle{unsrtnat}
%\bibliographystyle{plain}
%\bibliography{main}

\end{document}